\documentstyle[psfig]{mn}
\title[Spectroscopy of Extremely Red Galaxies in GOODS-South]{Deep GMOS
  Spectroscopy of Extremely Red Galaxies in GOODS-South: 
Ellipticals, Mergers and Red Spirals at $\bf 1<z<2$}
\author[N.D. Roche, J.S. Dunlop, K.I. Caputi, R. McLure, C. Willott, D. Crampton]
{Nathan D. Roche$^{1,2,5}$, James Dunlop$^{1,6}$, Karina I. Caputi$^{3,7}$,
Ross McLure$^{2,8}$\\ \\
 {\LARGE Chris J. Willott$^{4,9}$ and David Crampton$^{4,10}$}\\ \\
$^1$Institute for Astronomy,
     University of Edinburgh,
     Royal Observatory,
     Blackford Hill,
     Edinburgh EH9 3HJ,
     Scotland.\\
$^2$now at: Astronomy and Cosmology Research Unit, University of
 KwaZulu-Natal, Durban 4041, South Africa.\\
$^3$Institut d'Astrophysique Spatiale,
     bat.121 - Universite Paris XI
     91405 Orsay Cedex
     France.\\
$^4$Herzberg Institute of Astrophysics,
   National Research Council,
    5071 West Saanich Road,
    Victoria, B.C.,
    V9E 2E7,
     Canada.\\  
{$^5$ \verb"roche@ukzn.ac.za"}\hspace{8mm}
{$^6$ \verb"jsd@roe.ac.uk"}\hspace{8mm}
{$^7$ \verb"kcaputi@ias.u-psud.fr"}\hspace{8mm}
{$^8$ \verb"rjm@roe.ac.uk"}\hspace{8mm}
{$^9$ \verb"chris.willott@nrc.ca"}\hspace{8mm}\\
{$^{10}$ \verb"david.crampton@nrc.ca"}\hspace{8mm}
}

\begin{document}

\maketitle
\begin{abstract}

We have performed a deep (35.5 hours exposure) spectroscopic survey of
extremely red ($I-K>4$) galaxies (ERGs)
on the GOODS-South Field, using the GMOS spectrograph on the 8m
Gemini South Telescope.
We present here  spectra and redshifts for 16 ERGs
at $0.87<z<2.02$, to a limit of $K_s\simeq 20.2$.
In the spectra of  10 of these galaxies we find
 emission lines of [OII]$3727\rm \AA$,
with fluxes corresponding to a mean star formation rate (uncorrected for dust)
 of $1.6~ \rm M_{\odot} yr^{-1}$. For 3 ERGs
we detect no
 emission lines
      and the remaining 3
 lie at $z>1.5$ where this
line would be outside our wavelength range.
Absorption features are seen in most ERG spectra.

We examine the  morphologies of these ERGs on HST-ACS images and fit radii and
Sersic indices. We find
three broad classes:
(i) spheroidals; (ii) mergers at
 a variety
 of stages (some are spheroidal-disk pairs, some have tidal tails), often with a high surface brightness, and
 (iii) red spirals (which may have
star-forming regions in their outer disks).

We perform an  age-dating analysis by fitting the spectra and
9-band photometry ($BVIZJHK$, plus 3.6/4.5 $\mu \rm m$ fluxes from Spitzer)
 of the ERGs
with two-component models, consisting of passively
evolving, old stellar populations combined with a younger,
continuously star-forming component, for which the age
and dust extinction are allowed to vary from 10--800 Myr and $E(B-V)=0.0$ to 0.5 mag.
For only one ERG is the best-fit model purely passive, for
the others the best-fit is obtained by including  a star-forming component, which typically forms
  a few (0.26--13.5) percent of the stellar mass, and is subject
   to dust reddening averaging $E(B-V)\simeq 0.35$.
  The ages of the star-forming components tend to be
  youngest (10--40 Myr) in the merging ERGs, and
  older (200--800 Myr) in spiral ERGs, with mixed ages for the spheroidals.

The best-fitting
mean ages for the old stellar populations range from 0.6 to 4.5 Gyr,
 averaging 2.1 Gyr, with masses from
 $3\times 10^{10}$ to $2\times 10^{11} \rm M_{\odot}$.
The mean stellar formation redshifts of ERGs are spread from
$\sim 0.5$ Gyr  before
 the epoch of observation out to $z\sim 5$. We propose that
most ERGs are galaxies, or mergers of galaxies,
formed some Gyr earlier,
in an early ($z\sim 5$) phase of massive galaxy formation,
 which since then have experienced a wide variety of
merger and star-formation
 histories, accounting for the wide range of observed stellar ages.

Finally, we examine the clustering of the ERGs on this field, as a function
of the  photometric redshifts estimated by Caputi et al. (2004). The
comoving correlation radius is
 $r_{0}\simeq 13 h^{-1}$ Mpc for the full sample and, dividing by redshift,
  is constant or increasing with redshift, thus favouring
comoving ($\epsilon=-1.2$) evolution over a stable ($\epsilon=0$) clustering model.

 \end{abstract}
\section{Introduction}
The `extremely red galaxies' (ERGs, also known as EROs) are a
population of faint and distant galaxies, appearing in large numbers faintward
of $K\sim 18$,  with very red optical/near-infrared colours ($R-K>5$ or
$I-K>4$), corresponding to  galaxies which are already old and passively
evolving at $z>1$. The ERGs are of  great
importance in understanding galaxy evolution, in particular they represent
 an evolutionary link between the rapidly star-forming galaxies detected
by SCUBA at $z>2$, and present-day massive E/S0 galaxies.

A key property of ERGs is their strong clustering, as revealed by
measurements of their angular correlation function, $\omega(\theta)$
(e.g. Daddi et al. 2000; Firth et al. 2002;
Roche et al. 2002; Roche, Dunlop and Almaini 2003).
Assuming a comoving model of
clustering evolution, these observations give for ERGs
 a correlation radius $r_0\simeq 10$--$13 h^{-1}$ Mpc (where $h=H_0/100$ $\rm  km s^{-1}Mpc^{-1}$),
which supports their interpretation as the progenitors of giant ellipticals.

However, ERGs are actually a very diverse population of galaxies.
Spectroscopic observations find that some are
indeed old, passive
galaxies at $z>1$ (e.g. Dunlop 1996); however, at least half of the ERGs
 show emission lines
(Cimatti et al. 2002; Yan et al. 2004a)
 indicating ongoing star formation.
Radio, X-ray (Daddi et
 al. 2004) and mid-IR
(Yan et al. 2004b) observations indicate
 that many emission-line ERGs are
 actually undergoing major starbursts,
 with star-formation rates (SFRs) as high as $\sim 50$--100
$\rm M_{\odot}  yr^{-1}$.
Secondly, deep X-ray
surveys reveal at least $\sim 10$ per cent of ERGs
to host obscured AGN (Alexander et al. 2002; Roche et al. 2003). Thirdly, ERGs
are morphologically diverse; a mixture of bulges, disks, irregulars and
 mergers
 (e.g. Yan and Thompson 2003;
 Cimatti et al. 2003).

Initially we (Roche et al. 2002) colour-selected a
sample of 158 ERGs with $K\leq 21$ on the ELAIS N2 field. We found evidence of
strong clustering and  a mixture of  morphologies, and
identified 7 ERGs as  radio
sources (either AGN or powerful starbursts).
 We also fitted the ERG number counts with a model
in which they evolve into the present-day E/S0 population, through a
combination of passive luminosity evolution, merging,
 and a gradual increase with time in
the comoving number density of red galaxies.

For our second study (Roche, Dunlop and Almaini 2003, hereafter RDA03), deeper
 optical (HST-ACS) and near-IR (VLT-ISAAC)
data enabled us to  colour-select 198
 ERGs to a fainter limit
of $K=22$,
 on a central region of the Chandra/GOODS Deep Field South.
For the clustering of these ERGs we  estimated
$r_0=12.5\pm 1.2   h^{-1}$ Mpc, and we identified 17 ERGs with X-ray
sources in the 1-Msec
Chandra survey (Giacconi et al. 2002), from the X-ray fluxes
 diagnosing 13 as obscured AGN and the
 other 4  as probable starbursts.

In our third study, Caputi et al. (2004) fitted template spectra to the
7-band optical/near-IR photometry of these same ERGs to  derive
 photometric redshift estimates,  which in turn were used to estimate
stellar masses and
luminosity functions.  Caputi et al. (2005) then
extended this analysis to the full
sample of $K<22$ galaxies on this field (see Section 6).

In this paper we further investigate the CDFS ERGs by means of
 a deep optical spectroscopic survey, with  the
 Gemini Multi-object Spectrograph (GMOS).
We aim to  measure redshifts and emission lines for a sample of  ERGs
to a limit $K_{s}=20$--21, and furthermore to estimate the ages
of these galaxies by fitting
 evolving models to their spectra and broad-band fluxes (including new
3.6/4.5 $\rm \mu m$ fluxes from Spitzer).
We shall interpret and discuss these results and make comparisons with
the photometrically-based analysis of Caputi et al. (2004, 2005) and the
recent spectrographic surveys of  ERGs presented by
 Yan et al. (2004a), McCarthy et al. (2004), Daddi et al. (2004 and 2005)
    Doherty et al. (2005) and Longhetti et al. (2005).

In Section 2 of this paper we describe the optical/NIR data from RDA03 and
our new GMOS observations, and
in Section 3 the reduction and calibration of the
spectroscopic data. In Section 4 we present spectra for
the ERGs where we have secure
or probable redshifts, fit models to estimate ages
 and masses, and measure
 emission-line fluxes. In Section 5 we describe the morphology of the ERGs and
 estimate their size and surface brightness.
In Section 6 we re-examine the clustering of the RDA03
 ERGs as a function of  spectroscopic and
 photometric redshift, and in  Section 7 conclude with
interpretation and discussion.

Throughout this paper the assumed cosmology is $\Omega_m=0.25$,
$\Omega_\Lambda=0.75$, and $H_0=70$ km $\rm s^{-1}
Mpc^{-1}$, in accord with the recent (3 year WMAP) estimates of Spergel et al. (2006).
This gives the age of the Universe as 14.16 Gyr. Magnitudes are given
in the Vega system except where labelled by subscript as AB magnitudes,
 the conversion between the two being
$(K_s,H,J,I_{775})_{AB}= (K_s,H,J,I_{775})_{Vega}+(1.841, 1.373,
0.904, 0.403)$.
\section{Observations}
\subsection{Optical/NIR Imaging}
The Chandra Deep Field South forms
one of the two fields of the publically available
European Southern Observatories (ESO)  Great Observatories Origins Deep
Survey (GOODS). As an early part of GOODS, deep imaging in the
near-IR $H$, $J$ and $K_s$ bands was obtained
for a central region of the CDFS, covering
 50.4 $\rm arcmin^2$ and centered at RA $3^h 32^m 30^s$,
Dec. $-27:47:30$ (equinox 2000.0), using the
 ISAAC camera on the `Antu' Very Large Telescope (at the ESO site at Cerro
Paranal, Chile).
Also as part of GOODS, the central CDFS was mapped in four bands
 ($B_{435}V_{606}I_{775}Z_{850}$) with the
  Hubble Space Telescope Advanced Camera for Surveys
(ACS) (Giavalico et al. 2004). Most recently,
the GOODS Field South is currently being surveyed with Spitzer at
3.6--24 $\rm \mu m$. For this paper we have available 3.6 and 4.5 $\rm \mu m$
fluxes for the galaxies on this field,
derived from the `epoch 2' data release of May 2005.

Firstly, to define our ERG sample, we ran the
source detection program SExtractor on the mosaiced ISAAC $K_s$-band
data, to obtain a  sample
of $K_s\leq 22$ galaxies
(with completeness limit $K_s\simeq 21.5$). For more details
of this, see RDA03. With
 SExtractor we measured for these galaxies
 both total and aperture (2.0 arcsec diameter circular) magnitudes.

 These $K_s\leq 22$ galaxies were positionally
matched to the corresponding  detections on the ISAAC $H$ and $J$ mosaics
 and  the ACS $I_{775}$ images. Colours were measured for each as
 the differences in the 2.0
arcsec aperture magnitudes, and  the ERG sample was
 then selected as the $K_s<22$ galaxies redder than  $I_{775}-K_s=3.92$, which
numbered 198 in the 50.4 $\rm arcmin^2$ area.
We numbered these objects 1 to 2555 using their order of detection in the $K_{s}<22$
 catalog.

For the galaxies studied in this paper
 we also measure 3.6 and 4.5 $\rm \mu m$ fluxes from
the Spitzer `epoch 2' data release of May 2005.
The Spitzer data has a lower resolution of $\rm FWHM\simeq 1.6$ arcsec.
           Because of this, larger apertures
of 2.8 arcsec diameter
were employed to measure the Spitzer fluxes, and to these fluxes
aperture corrections of 0.5 mag for $3.6 \rm \mu m$ and 0.55
mag for  $4.5 \rm \mu m$ were applied, as derived from the point-spread
function.

\subsection{Spectroscopic Observations}
The target galaxies for our GMOS survey form a subset of the RDA03 sample,
consisting primarily of the brighter,  $K_{s}<21$, ERGs within
 a  $5.5\times 5.5$
arcmin area that can be covered by a single GMOS mask.
A spectrograph mask was prepared with 47 slitlets
 (each of length 2.0 arcsec and width 1.0 arcsec) devoted to our target ERGs.
Figure 1 shows their positions
within the larger area of our CDFS sample.
\begin{figure}
\psfig{file=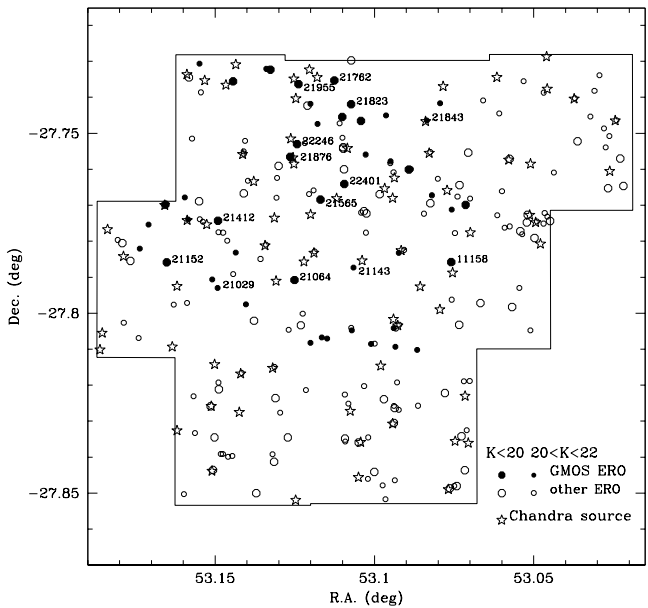,width=90mm}
\caption{Positions of the 198 ERGs in the CDFS sample of RDA03,
 within the borders of its 50.4 $\rm arcmin^2$
 area. The {\it Chandra} X-ray sources (Giacconi et al. 2002) are also shown.
 Bold symbols indicate the 47 ERGs targeted by our GMOS spectroscopic survey,
and the 16 of these which form
our redshift sample (see Section 3 and Table 1) are labelled
 with their ID numbers.}
\end{figure}
Observations were performed with GMOS on the Gemini South telescope in Chile,
 over the period from 24
September 2003 to 21 February 2004, using the low-dispersion, R150
grating and central wavelengths in the region of
$\rm 7500\AA$.
 In total we were able to obtain 70 exposures,
totalling 35.5 hours, through the ERG slit mask.
These science observations were interspersed with CuAr
arc lamp and standard star exposures, for wavelength and
flux calibration.

These observations utilized GMOS in the the `nod-and-shuffle' mode
described in detail by Abraham et al. (2004).
The sky background is a serious problem
for ground-based spectroscopy, especially  at redder, $\lambda>7500\rm \AA$,
wavelengths where there are strong emission lines.
The `nod-and-shuffle' mode was designed to provide improved
subtraction of the sky background, even where this is subject to
considerable spatial and time variation.
Essentially, the
telescope is `nodded' (by 1 arcsec in this case) along the spectroscopic slit,
observing at each position alternately
 in a rapid one-minute cycle. In position A the object will be towards
 one side of
 the slit and the adjacent sky on the other, in position B the object will be
 on the other side.
With each cycle, the charges stored on
the CCD from the two positions
 are `shuffled' (by 2 arcsec) into adjacent
strips on the CCD frame, where they are accumulated over the course of
each half-hour exposure. In this way the sky is observed simultaneously with,
and as close as possible in position to, each individual target galaxy.

\section{Data Reduction}
GMOS data reduction was carried out using {\sevensize IRAF} routines,
including some specifically written for this instrument.
Firstly, bias frames were subtracted from the  target galaxy, arc-lamp
 and standard star exposures. The next step was sky subtraction, performed
 individually for each of the 70 exposures.
The nod-and-shuffle mode produces a CCD frame showing two adjacent
strips for each spectrograph slit, corresponding to the two nod positions.
To sky subtract, the frame is shifted by
the shuffle step and subtracted it from itself (using the
{\sevensize IRAF} `gnsskysub' task). This produces a positive and a negative
sky-subtracted spectrum for each target galaxy.

Small dithers in the spectral and spatial directions between the 70
exposures (introduced intentionally for the removal of charge traps; Abraham
et al. 2004) were removed by
 registering the 70 exposures in both axes.
The 70 sky-subtracted frames could then be combined, using {\sevensize IRAF}
`imcombine' with  a `sigclip' rejection to remove most of the cosmic rays,
into a single 35.5
hour image, from which we extract the spectra.

{\sevensize
IRAF} tasks (e.g. `apall') were then employed to extract the
2D spectra from each slit,
summing each in the spatial direction to give a 1D spectrum.
 With nod and shuffle it is necessary to extract
both the positive and a negative spectrum for each object,
and then subtract one from the other.

 The galaxy
 spectra were wavelength calibrated using
   the CuAr lamp calibration spectrum as observed through the same slit
 of the mask. Each pixel is approximately $3.5\rm \AA$
 and the resolution of our spectra $\rm FWHM\simeq 18\AA$
Flux calibration was performed using a co-added spectrum of the calibration
 standard star LTT1020, and  the spectral energy distribution of this star as
tabulated by Hamuy et al. (1994), to            derive
(using {\sevensize IRAF} `sensfunc') a sensitivity function
for the spectrograph (Fig 2a).

Corrections were also applied to the spectra
for the telluric (Earth's atmosphere) absorption bands,
 using our calibration
spectrum of a white-dwarf standard star (these having few intrinsic spectral features), EG131. This was
divided  by a fit to its own continuum, giving a function
almost flat at unity
except for four telluric absorption bands, centred at 6560,
6875, 7610 and $9075\rm \AA$. These were extracted,
converted into magnitude units, and inserted into a function
$y(\lambda)=1$ to give a telluric correction function.

 The ERG spectra were divided by this correction function
and by the sensitivity function, to
give spectra calibrated in $F_{\nu}$ (in units of $10^{-29}$ ergs
$\rm cm^{-2}s^{-1} Hz^{-1}$, i.e. $\rm \mu Jy$),
and $F_{\lambda}$ (in units $10^{-19}$ ergs
$\rm cm^{-2}s^{-1} \AA^{-1}$), ready for further analysis.
\begin{figure}
\psfig{file=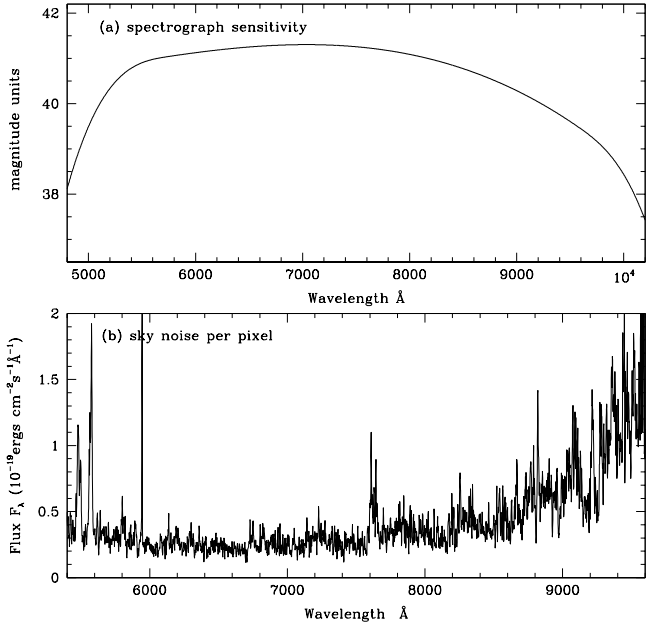,width=90mm}
\caption{(a) Sensitivity of the spectrograph as a function of wavelength
(as measured using calibration star
 LTT1020), and (b) sky noise within an
extracted spectrum, calibrated in $F_{\lambda}$.}
\end{figure}
\section{Spectra of the ERGs}
\subsection{Defining the Redshift Sample}
On examining the GMOS spectra we found that our survey had a good rate of
success for ERGs to  $K_{s}\simeq 20$, but failed to reach the fainter
$K_{s}=21$ limit of the target list.
Of our 47 spectra of ERG targets, 16 were galaxy spectra of sufficient
 quality that we were
able to determine good or, at least, likely redshifts.
17 showed no significant
signal, 13 had some signal in the continuum but we were unable
to confidently
identify any spectral features above the noise, and one
was identified as a red
Galactic star.
The 16 `successful' spectra
include 13 of the 18 target ERGs with $K_{s}<20$, 3 of the 6 target
ERGs with
$20<K_{s}<20.25$ but none fainter than this.

Redshifts were estimated by a combination of (i) fitting models to the
continuum, by the method described below, (ii) identifying absorption features,
and (iii) from emission lines such as
 [OII] $\rm 3727\AA$, where present. For galaxies with an
emission line we determine both an emission ($z_{em}$) and a
continnum-fit/absorption-line
($z_{cf}$) redshift. In all cases we found the two consistent within the
$\sim 0.01z$ statistical error on $z_{cf}$. Where both are given
$z_{em}$ is likely to be more precise.

 Two of these redshifts, for the galaxies 1843 and
1158, are based on spectral features of poor signal-to-noise and must be
regarded as less reliable than the other 14. However these two redshifts are
supported by very good fits of our model spectra to the 7-band photometry.

We therefore define a redshift sample comprised of these 16 ERGs.
The 16  redshifts range from $z=0.0868$ to $z=2.017$, with
 6 in a `spike' at $1.180\leq z\leq 1.225$, which may be
indicative of large-scale structure.
One of the 16 ERGs, ID no. 1843, corresponds to a {\it Chandra} source.
This is XID:253 in the list of Giacconi et al. (2002), with a flux and
hardness ratio ($HR=0.66\pm 0.11$) indicative of an
obscured AGN (see RDA03).

\subsection{Comparison with Spectroscopy of Other Authors}
Some of the objects in our redshift sample were recently observed
spectroscopically by other authors, in separate and independent studies,
published
  during the writing of this paper -- Vanzella et al. (2005), who used FORS2 on the
  VLT, and Doherty et al. (2005), using DEIMOS on Keck. For objects in the overlap
   between these surveys, we make a brief comparison of the spectra.

\noindent (i) Our object 1152 is in Vanzella et al. (2005) as
GDS J033239.64-274709.1. The FORS2 spectrum gave $z=1.317$, consistent with
our $z=1.309$, and similarly shows weak [OII] emission,
 and absorption lines. 1152 is also object 2800 in Doherty et al. (2005), observed
with DEIMOS giving $z=1.318$, with $\rm H\&K$ absorption..
 Neither spectrum goes far enough into the red to see well the Balmer lines.

\noindent (ii) Object 1029 was observed as GDS J033235.79-274734.7. The FORS2
spectrum gave $z=1.223$, consistent with our $z=1.221$, and is very
 similar to ours, with strong [OII] emission and some absorption
  features, including (weak) Balmer lines.

\noindent(iii) Object 1412 is GDS J033235.78-274627.5. The FORS2
spectrum gave $z=1.094$, consistent with our $z=1.092$, and like ours shows [OII]
emission. The H8 and H9 lines we find are not identified on the FORS2 spectrum,
 but it
 extends further into the red to reveal $H\gamma$ and $H\delta$ absorption.
  Thus on the basis of both spectra this seems to be a strong Balmer
   absorption galaxy.

\noindent(iv) Object 1777 was not observed in the FORS2 survey.

\noindent(v) Object 2536 is GDS J033231.83-274356.2. The FORS2 spectrum
 gave $z=1.550$, slightly higher than our $z=1.513$. Like our GMOS spectrum it
 features MgII ($2800\rm \AA$)
 absorption, and extends a little further redwards, detecting an [OII]
 emission line.

\noindent(vi) Object 1064 is GDS J033230.02-274726.8.
  The FORS2 spectrum gave $z=1.553$,
 a close match to our $z=1.548$, and detects MgII ($2800\rm \AA$)
 absorption, and a weak [OII] emission line at the reddest extreme,

\noindent(vii/viii/ix) Objects 1143 and 1565 and 2401 were not observed in the FORS2 survey.

\noindent(x) Object 1876 is GDS J033230.34-274523.6. The FORS2 spectrum gave $z=1.223$,
slightly higher than our $z=1.196$. The FORS2 spectrum, like ours, shows MgII ($2800\rm \AA$)
absorption, [OII] emission and Balmer lines, extending further redwards to $H\gamma$.
1876 is also number 2158 in the Doherty et al. (2005) DEIMOS sample, where it has
$z=1.222$ and $\rm H\&K$, $\rm H\delta$ absorption, very weak [OII], and is classed as $\rm E+e$

\noindent(xi) Object 2246 is  GDS J033229.82-274510.8. The FORS2 spectrum gave $z=1.225$,
slightly greater than our $z=1.196$. Like our
 spectrum it shows [OII] emission and $\rm H\&K$ absorption, the spectra are very similar but
  ours has better signal-to-noise.

\noindent(xii/xiii) Objects 1823 and 1955 were not observed in the FORS2 survey.

\noindent(xiv) Object 1762 is GDS J033227.02-274407.2. The FORS2 spectrum gave $z=1.127$,
 slightly higher than our $z=1.102$, and like ours shows
  moderate [OII] emission and weak Balmer absorption. The authors mark $H\gamma$ but
   it is barely visible.

\noindent(xv) Objects 1843 and 1158  were not observed in the FORS2 survey. However, the AGN 1843
 was observed with VLT FORS in an earlier survey of Chandra sources. Szokoly et al. (2004)
 initially proposed $z=0.484$ on the basis of a single unidentified line.
 Zheng et al. (2004) claimed $z=1.89$ on the basis of a single line, this time supported by
 photometric redshifts from COMBO-17 and hyper-$z$. However, for our 9-band photometry we
 obtain a hyper-$z$ solution of $z=2.06$, and our continuum is significantly better fitted for
 $z=2.017$ than $z=1.89$. Considering that photometric redshifts are uncertain by at least
 $8\%$, we retain $z=2.017$ as a provisional and approximate estimate. Near infra-red spectroscopy should
 detect multiple emission lines and give an unambiguous answer.
\smallskip

In summary, out of our redshift sample of 16, 8 objects were also observed by
Vanzella et al. (2005) and/or Doherty et al. (2005),
 who obtained very similar spectra  similar spectra (although in two cases with a redshift
discrepancy that might be larger than expected from statistical error). Seven  of our redshifts
appear to be  new. Finally, for the $z\sim 2$ AGN, more data is needed.
\subsection{Model-fitting}
We fit our ERG spectra, at $5800<\lambda<9000\rm \AA$
where  the signal-to-noise is greatest, with a series of
40 template spectra from Jimenez et al. (2004).
These represent passively evolving stellar
populations, with a Salpeter IMF and solar metallicity, at ages from 0.001
to 14 Gyr. We experimented with models of lower and higher metallicity but
found no indication that these fitted any better. Furthermore the colours
(Caputi et al. 2004) of our ERGs and the ultraviolet spectroscopy of
de Mello et al. (2004) support the assumption that ERGs have
 approximately solar
metallicity, even to $z\sim 2$.

From the scatter between the 17 `blank' (i.e. no signal) slit spectra,
we derived a
function representing the noise per pixel produced by
 the sky background within the area of a
single spectroscopic slit (Fig 2b). This is much worse at the red end due to
strong sky emission lines and the declining detector response.

We fit models to a combination of the GMOS spectra and the 9-band ($BVIZJHK$,
$3.6/4.5\mu \rm m$) photometry, using
a least-squares (minimum
$\chi^2$) method,
 with each pixel of the spectrum weighted
using this noise function. The
GMOS spectrum and the 9-band photometry make similar contributions to
to the combined $\chi^2$, which is minimized in the fitting.

An initial series of fits as a function of redshift
provided the continuum-fit redshift $z_{cf}$ for
each ERG. We then
 fixed the redshift at this value and  fit with a   {\it two-component} model,
this being the  sum of a single-age
passively evolving model, of age $T_{pas}$, and
 a `starburst' component of much younger stars, represented by the
Jimenez et al. (2004) model, integrated with a constant SFR for a time interval $T_{sb}$
prior to the time of observation. It is important to note that this component is not a
single-age population, but contains stars
 of all ages from zero to $T_{sb}$. The young component is subject to
 dust-reddening, following the extinction curve of  Calzetti et al. (2000),
 parameterized in terms of $E(B-V)$.
The normalization of the young component is parameterized as $f_{sb}$ the fraction of the
emitted flux it contributes in the rest-frame $B$-band, $4500\rm \AA$. Hence it is described by three
parameters, $T_{sb}$, $E(B-V)$ and $f_{sb}$.

For each ERG, we searched for the best-fitting model, by calculating a $\chi^2$
for the GMOS spectrum,
plus the 9-band
 photometry compared with a set of model spectra,
 forming a grid of 18000 points in the 4-dimensional ($T_{pas}$, $f_{sb}$, $T_{sb}$,
 $E(B-V)$) parameter space. $T_{sb}$ was allowed to vary from 10 to 800 Myr and $E(B-V)$
 from 0 to 0.5 mag ($A_V\sim 2$ mag).
The best-fit model was taken as the position
  within this grid where $\chi^2$ has the minimum value $\chi^2_{min}$.

  Error intervals were estimated for the fitted parameters, as follows. Using our grid
  of models and their associated $\chi^2$, we  find the surface in parameter space,
   enclosing the  best-fit model,
   on which $\chi^2$ takes the value $\chi^2_{min}+2.71 (\chi^2_{min}/N_{df})$,
    where $N_{df}$ is the number of degrees of freedom, approximately 900 here.
     This surface defines
    an approximately $90\%$ confidence interval.
    The error on a given parameter, e.g. $T_{pas}$, is then taken as
    the {\it projection} of this entire
    surface onto the $T_{pas}$ axis ($not$ just its intercept with the axis).
    This projected interval will then
    take into account any degeneracies between the fitted parameters,
     which would produce an elongated surface and hence widen the
     projected error interval.

The Jimenez et al. (2004) models are normalized in units of solar masses  and
so the normalizations of the model fits can provide estimates of the stellar
mass in each galaxy, for both the passive ($M_{pas}$) and the young ($M_{sb}$)
stellar components.
 However, the model fits to the 9-band aperture photometry
always gave higher normalizations than fits to the spectra (by
factors 1.1--2.5), the reason being that  a 2.0 arcsec
diameter aperture admits a greater fraction
 of the light from an extended galaxy than
the 1.0 arcsec spectrograph slit. We therefore  give stellar
mass estimates with the higher normalization
given by fitting models to the aperture
photometry.

\subsection{Spectra and Ages of the Redshift Sample}
Figure 3 shows plots of our 16 GMOS spectra, the
 9-band photometry and best-fit models. Spectra and models are normalized to
 fit the aperture (rather than slit) photometry.

 Table 1 lists co-ordinates of the 16 galaxies and Table 2 the best-fit parameters
$T_{pas}$, $M_{pas}$, $M_{sb}$ and $f_{sb}$, with approximately 90 per cent confidence
interval error bars.
 Also given are a flux-weighted mean star-formation redshift,
$z_{msf}$, which corresponds to the look-back time $T_{pas}$ earlier than the
redshift of observation, the masses of the passive and starburst components
in the best-fit model, plus $Mf_{sb}(={M_{sb}\over M_{pas}+M_{sb}})$
the fraction of the total stellar mass in the starburst/young
component.
 Table 3 gives the 9-band aperture photometry,
 absolute magnitude $M_B$ (AB system) in the rest-frame $B$-band,
  derived from the best-fit model normalized to the 2.0 arcsec aperture magnitudes,
  and $M_{BT}$ an absolute magnitude with an approximate correction to total magnitude
  given by adding the (small) difference between
aperture and total $K$ magnitudes.

We find  a wide variation in $T_{pas}$ within the sample,
  from 0.6
Gyr to 4.5 Gyr, with a mean of  $2.1\pm 0.3$ Gyr, and in the mass $M_{pas}$,
  from $3\times 10^{10}\rm M_{\odot}$ to almost $2\times 10^{11}\rm M_{\odot}$,
  with a mean $1.1\times 10^{11}M_{\odot}$.
Note that we have performed
two independent reductions of the GMOS data, and the
other reduction was independently fit with both Jimenez et al. (2004) and
Bruzual and Charlot (2003) models. These  analyses gave age estimates closely
 similar to, and consistent with, those presented here.

 Only for one ERG was the best-fit obtained for a purely passive model,
$f_{sb}=0$. For the 15 where $\chi^2$ is smaller  with $f_{sb}>0$,
the best-fit star-forming component made up
 anything from 0.26 to 13.5 per cent of the total stellar mass
 (mean $3.5\pm 1.1$ per cent)
The best-fit ages and dust-reddenings for the star-forming
 components vary widely between
individual ERGs, and  cover the full range allowed in this model,
10--800 Myr (mean $364\pm 96$ Myr) and 0-0.5 mag (mean $0.35\pm 0.05$).

For 6 ERGs, the best-fit star-forming component is a short-term,
$<100 $ Myr starburst, always with some dust reddening varying from $E(B-V)=0.2$ to 0.5.
 For 9 others, the best-fit is with
a longer period (100--800 Myr) of continuous star-formation, in
3 cases with little/no dust but
in the other 6 with heavy extinction of $E(B-V)\simeq 0.5$.
In Section  7.3 we discuss how age and recent SF history is related to morphology.

\subsection{Spectral Lines}
Ten of the 16 ERG spectra show  a single emission line, which in all cases
is assumed to be  [OII]$3727\rm \AA$. In most cases this
identification is supported by the position of other spectral features.
 [OII]$3727\rm \AA$ emission is an indicator of ongoing or
very recent (within $\sim 10^7$ yr) star-formation, and according to Kennicutt
(1998), a luminosity $L_{\rm OII}$ ergs  $\rm s^{-1}$ in this line corresponds to
a SFR $1.4\times 10^{-41}L_{\rm OII}$ $\rm M_{\odot}yr^{-1}$.

 For 3 ERGs we find no emission at
$3727\rm \AA$, indicating that star-formation has ceased, is
at a very low level, and/or is heavily obscured. One of these (2401) has a spectrum
fit by a purely passive model, but the photometry of the other
 two (1701 and 1955) is best-fit
with the inclusion of a heavily reddened young component,
suggesting these are dusty post-starbursts.
Three ERGs
lie at $z>1.5$ where [OII] would not be
visible in our wavelength range, but two of these have [OII] visible
at the red extreme of the
FORS2 spectra (which have a slightly redder limit), and all three
have strong fluxes at the blue end.

Absorption features are prominent
 in almost all ERG spectra,
the MgII close doublet at $\rm 2800\AA$ and/or $\rm H+K$ (3934, $3968\rm \AA$),
 and  the $\sim \rm 4000\AA$ break is generally a prominent feature.indicating the presence of old stellar populations.
Three emission-line ERGs (1412, 1823, 1876) show H8 and H9 Balmer absorption lines
 indicating a large  intermediate-age
 ($10^8$--$10^9$ yr) component.

The [OII]$3727\rm \AA$ line fluxes  were measured from the GMOS spectra
(using {\sevensize IRAF} `splot'), and multiplied by the ratio of the
continuum flux in the 2.0 arcsec aperture to that in the GMOS slit,
 so as obtain a line flux more representative of the entire galaxy.
The [OII] line fluxes are listed in Table 4, with errors calculated by summing
the sky noise function (Figure 2b) over the resolution element ($18\rm \AA$)
centred on the observed line wavelength. From this flux and the spectroscopic
redshift we derive the  [OII]$3727\rm \AA$ luminosity of the galaxy,
 which is uncorrected for dust, and from this a star-formation rate $\rm SFR_{OII}$,
 using the relation of Kennicutt (1988).

These SFRs, which again are uncorrected for dust extinction,
average $1.61\pm 0.37 \rm ~M_{\odot}yr^{-1}$. There are two ways
in which we can estimate a very approximate dust-corrected SFR.
Firstly, if the emission lines are assumed to be subject to the same dust extinction
  as the continuum,  then from the Calzetti (2000) reddening
 curve the [OII] extinction $A_{3727}=5.86E(B-V)$. If we apply this correction to the SFRs in
 Table 4, using the value of $E(B-V)$ best-fit to the spectrum of each galaxy, the mean
SFR of the 10 emission-line galaxies increases to $12.6 \rm ~M_{\odot}yr^{-1}$

Secondly, if we
assume that the SFR and dust extinction in the star-forming component is truly constant, we can
estimate $\rm SFR_{sb}=M_{sb}/T_{sb}$. This will correct for dust, as this is taken into account
 in calculating $M_{sb}$, but for an
individual ERG this is very approximate due to the large error bars on
$T_{sb}$. For the full sample of ERGs (but excluding the AGN), the mean $\rm SFR_{sb}$ is
$24 \rm ~M_{\odot}yr^{-1}$.
\begin{table}
\caption{RA and Dec co-ordinates of the 16 ERGs in our redshift sample; and
 total K magnitude (Vega system)}
 \begin{tabular}{lccc}
 \hline

ID no. & R.A. & Dec. & $K_{total}$\\
\smallskip
  &   (deg.) & (deg.) \\
1152 &  53.165199 & -27.785870 & 19.14 \\
1029 &  53.149200 & -27.792976 & 20.21 \\
1412 &  53.149109 & -27.774279 & 19.23 \\
1777 &  53.144367 & -27.735580 & 18.87 \\
2536 &  53.132652 & -27.732313 & 19.60 \\
1064 &  53.125080 & -27.790777 & 18.98 \\
1143 &  53.106606 & -27.787319 & 20.23 \\
1565 &  53.116936 & -27.768448 & 19.77 \\
2401 &  53.109440 & -27.764095 & 18.93 \\
1876 &  53.126465 & -27.756538 & 18.24 \\
2246 &  53.124275 & -27.752996 & 19.68 \\
1823 &  53.107304 & -27.741930 & 18.61 \\
1762 &  53.112560 & -27.735302 & 18.66 \\
1955 &  53.123837 & -27.736311 & 18.91 \\
1843 &  53.083546 & -27.746445 & 20.00 \\
1158 &  53.075893 & -27.785822 & 19.78 \\
\hline
\end{tabular}
\end{table}

\newpage
\begin{figure}
\psfig{file=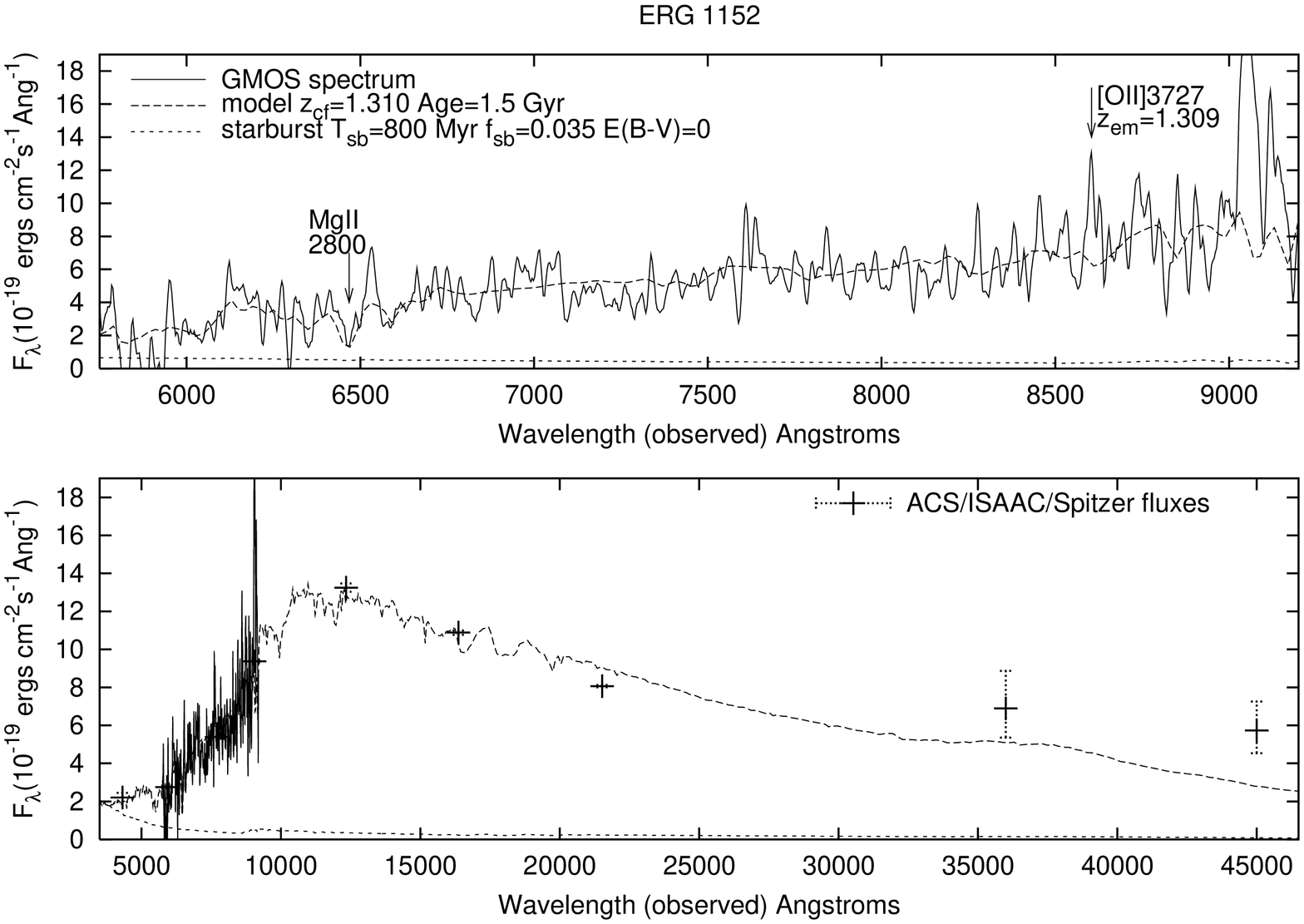,width=90mm,angle=-90}
\end{figure}
\begin{figure}
\psfig{file=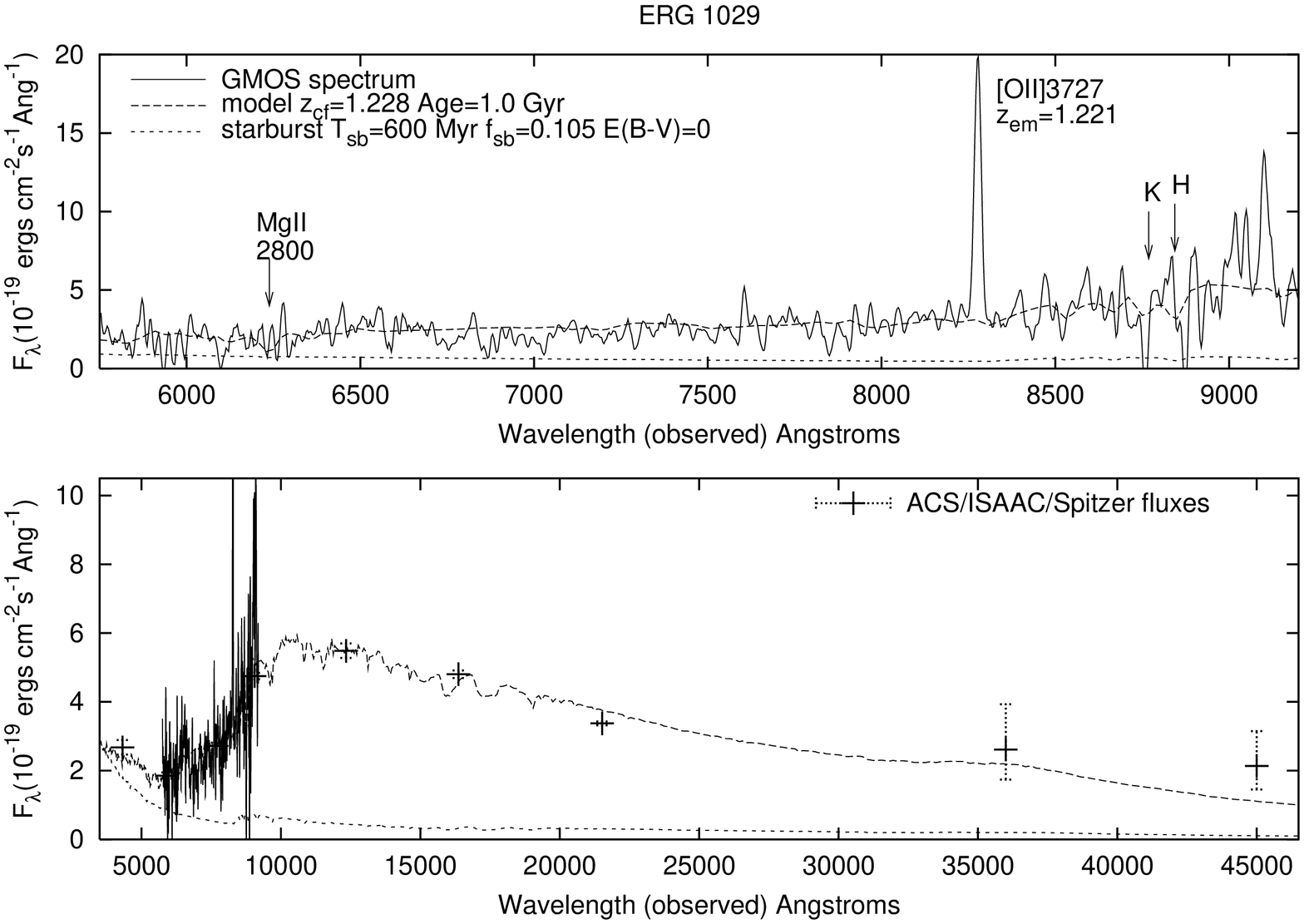,width=90mm,angle=-90}
\end{figure}
\begin{figure}
\psfig{file=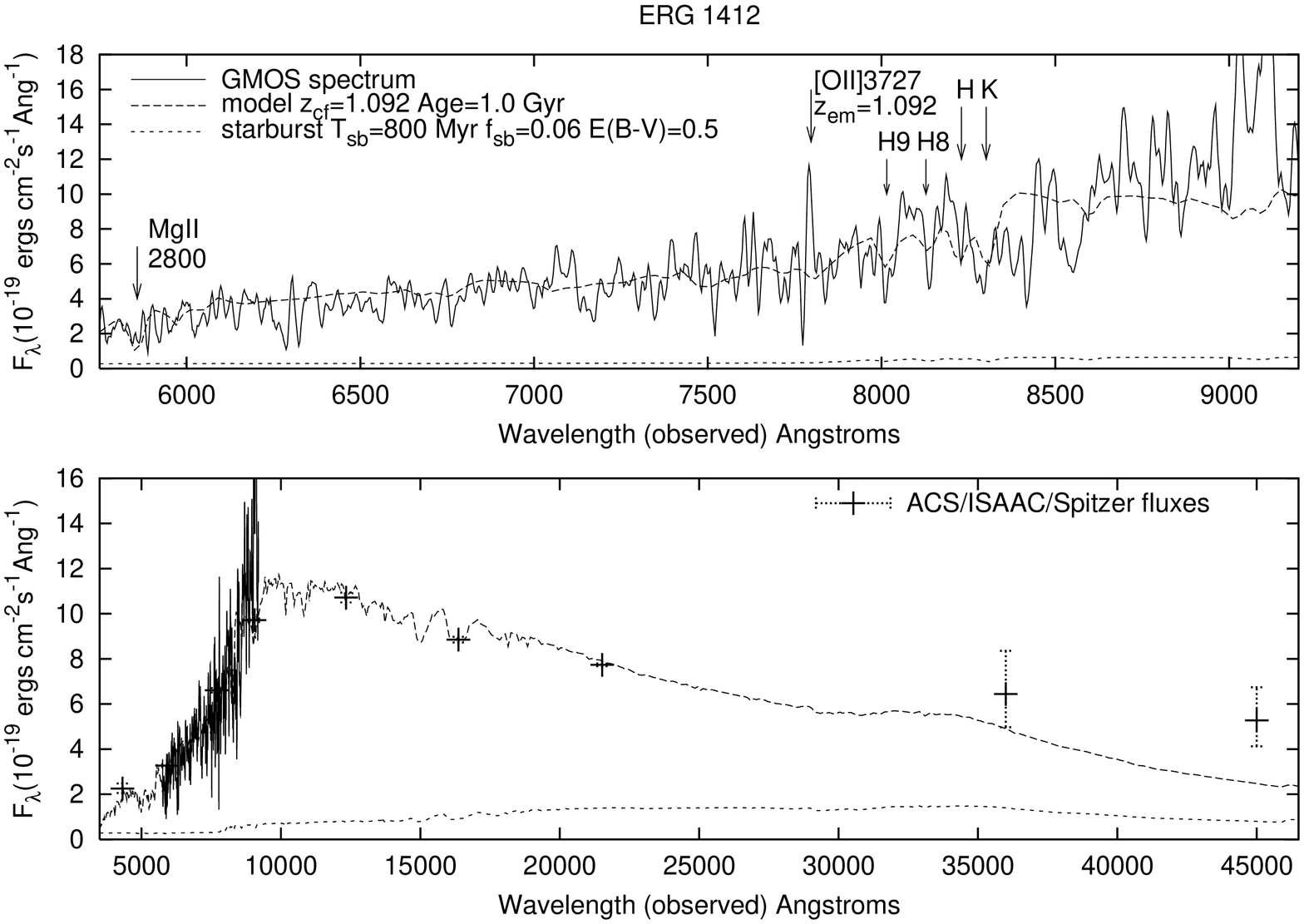,width=90mm,angle=-90}
\end{figure}
\begin{figure}
\psfig{file=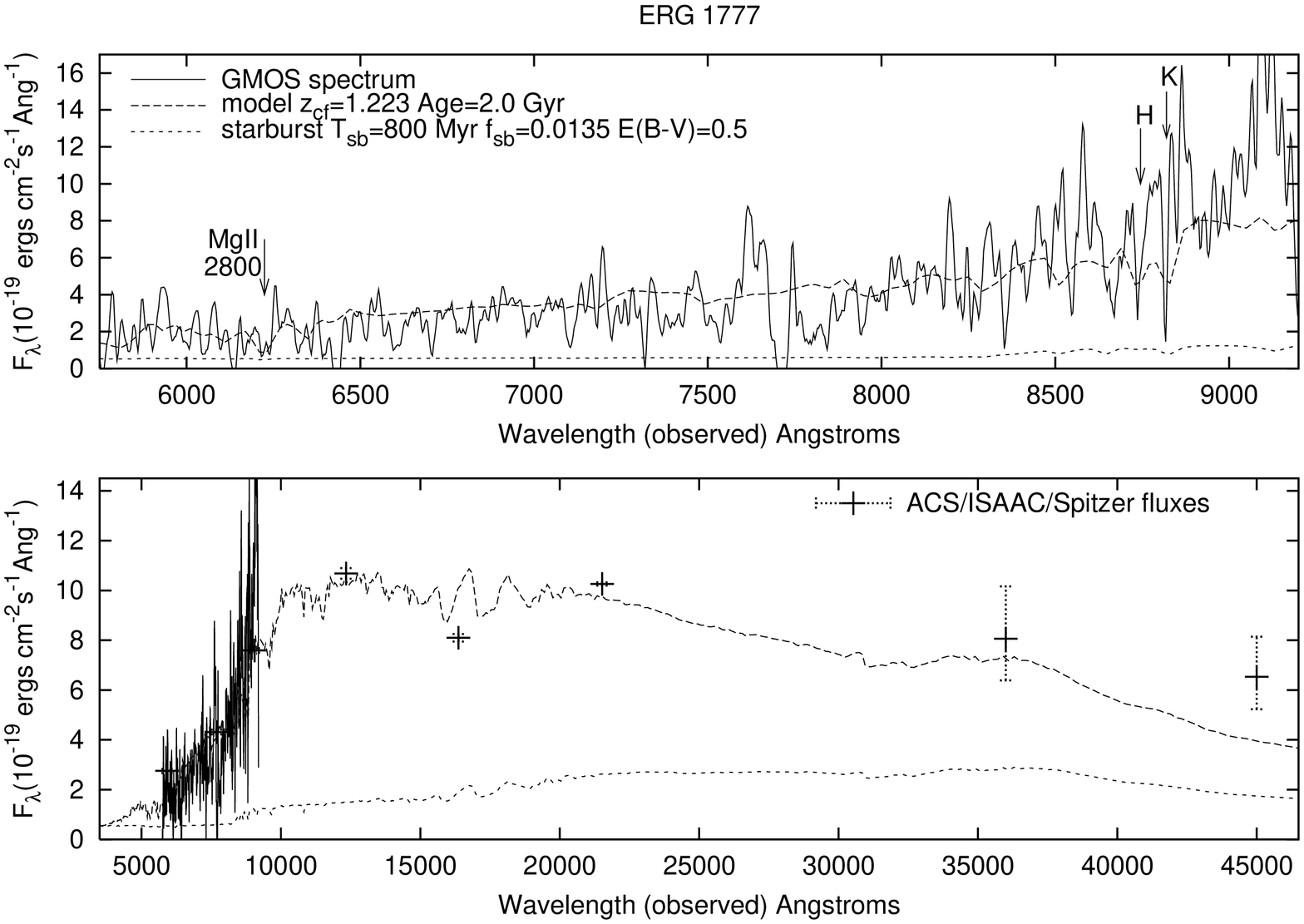,width=90mm,angle=-90}
\end{figure}
\begin{figure}
\psfig{file=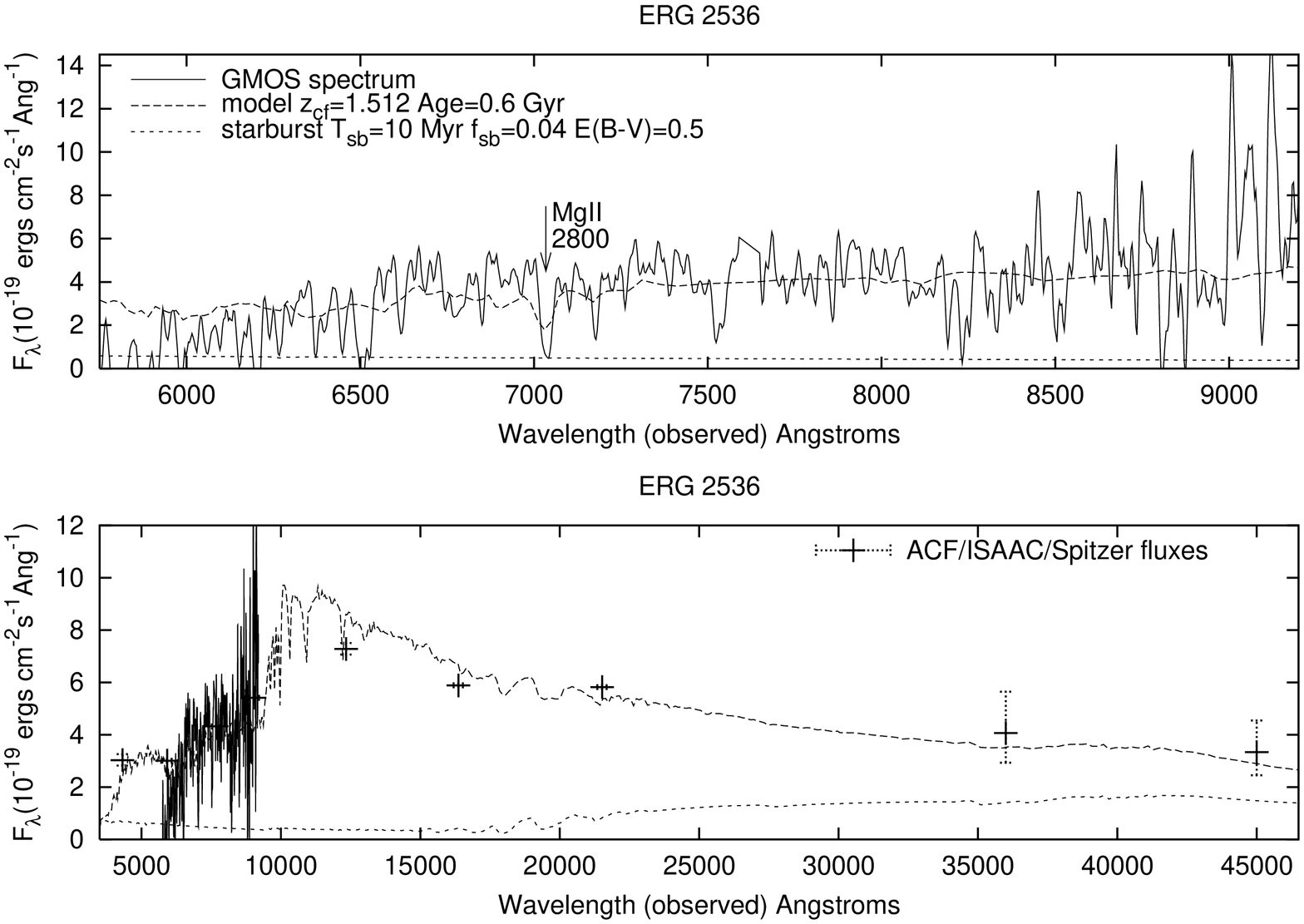,width=90mm,angle=-90}
\end{figure}
\begin{figure}
\psfig{file=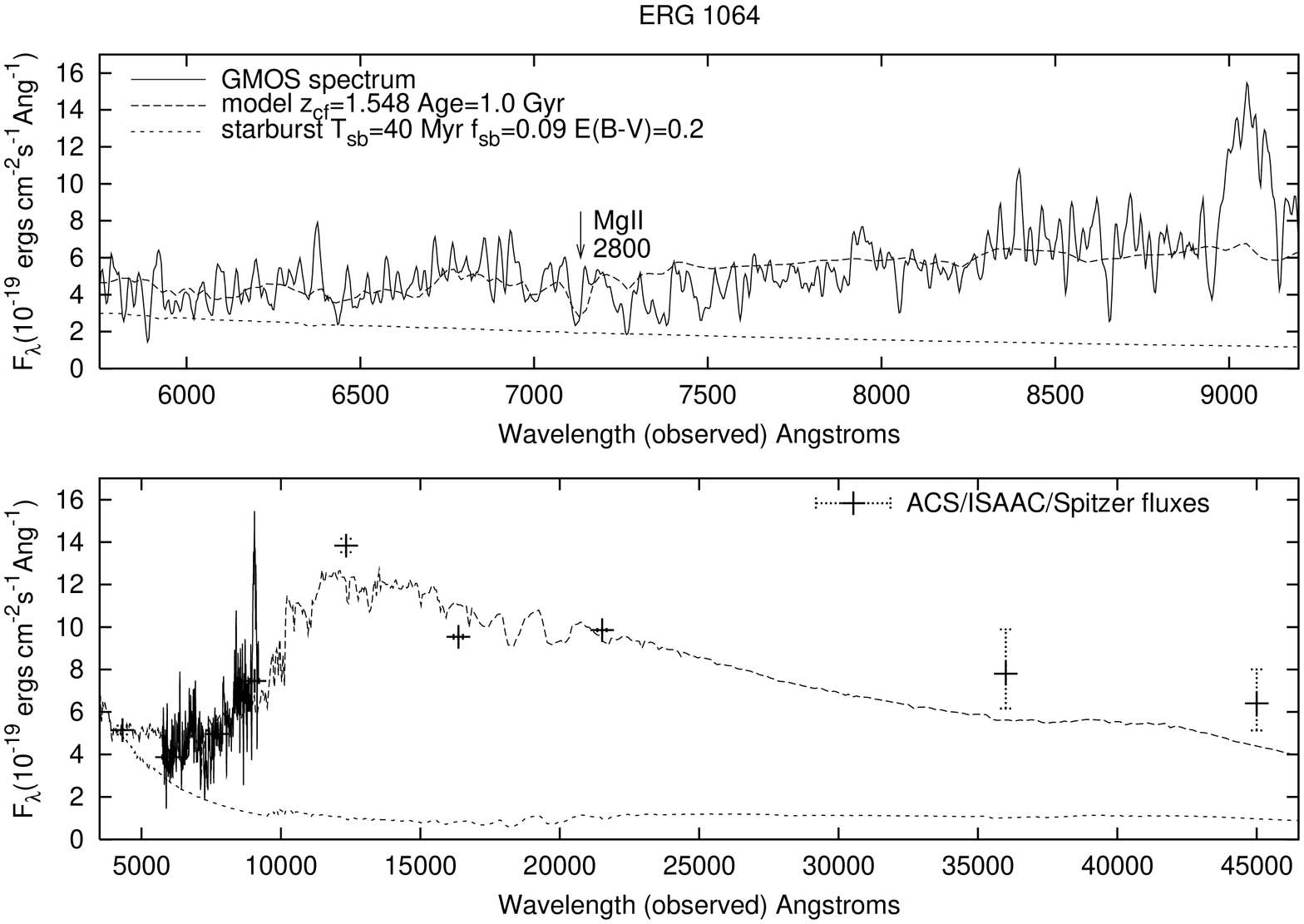,width=90mm,angle=-90}
\end{figure}
\begin{figure}
\psfig{file=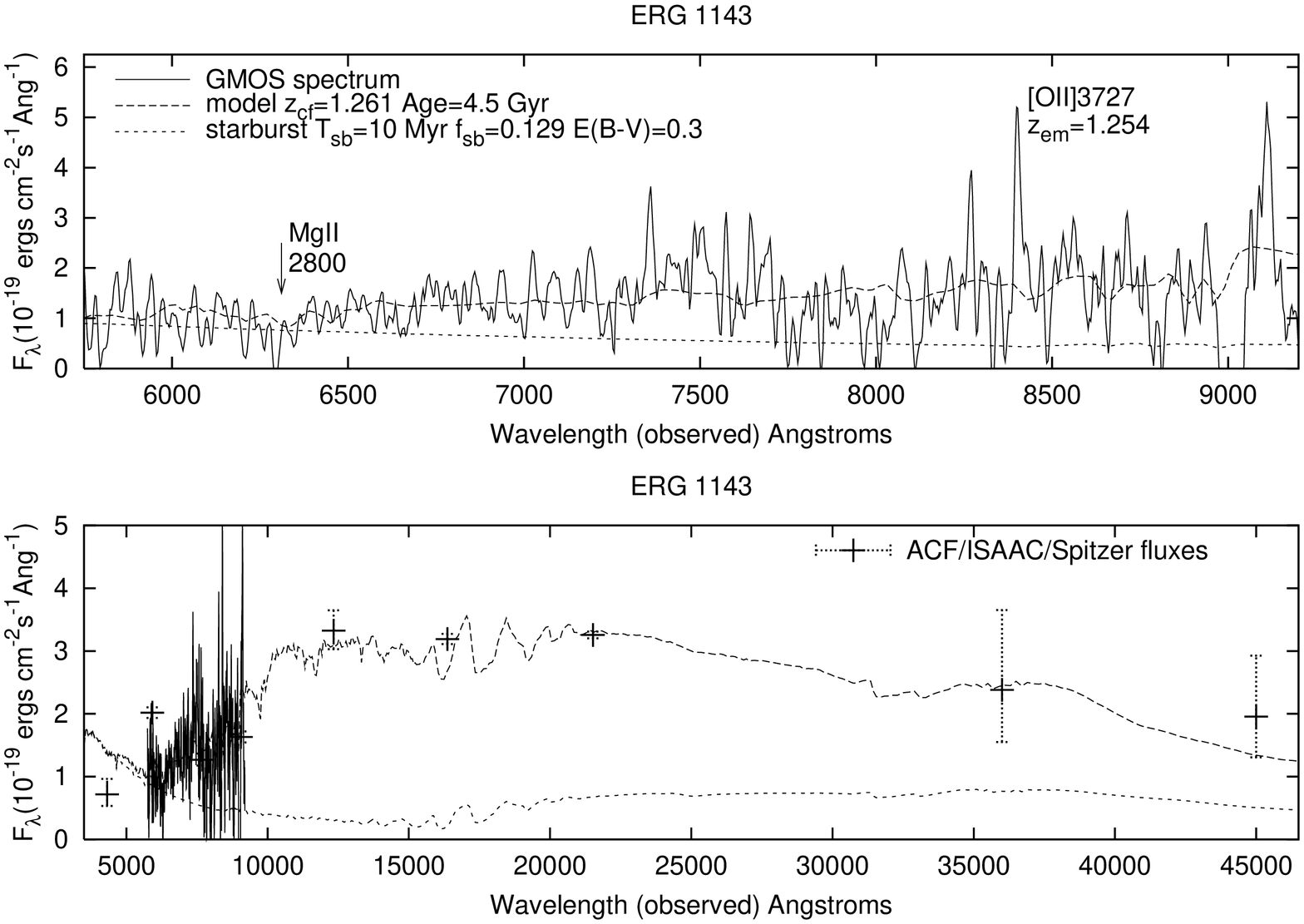,width=90mm,angle=-90}
\end{figure}
\begin{figure}
\psfig{file=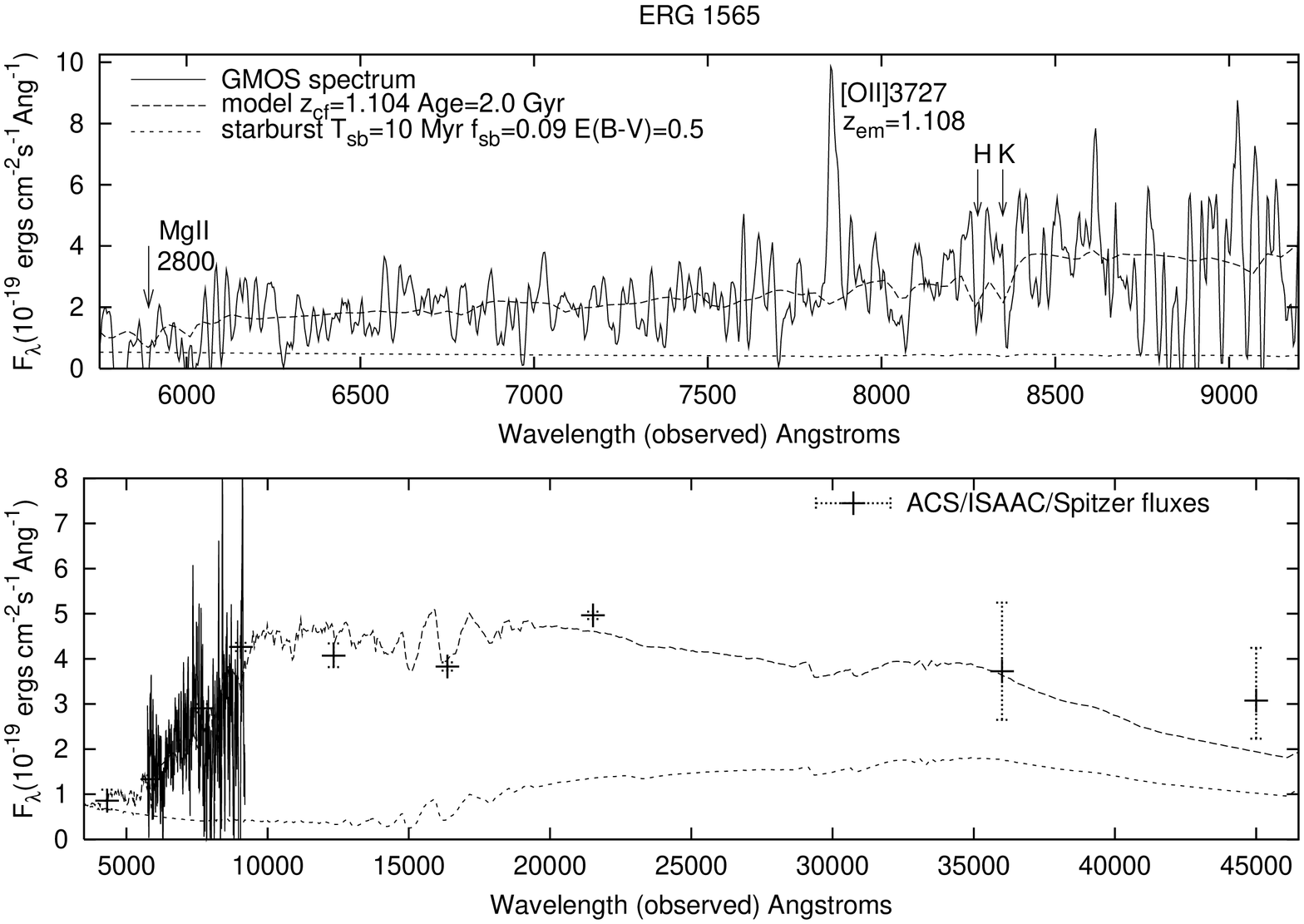,width=90mm,angle=-90}
\end{figure}
\begin{figure}
\psfig{file=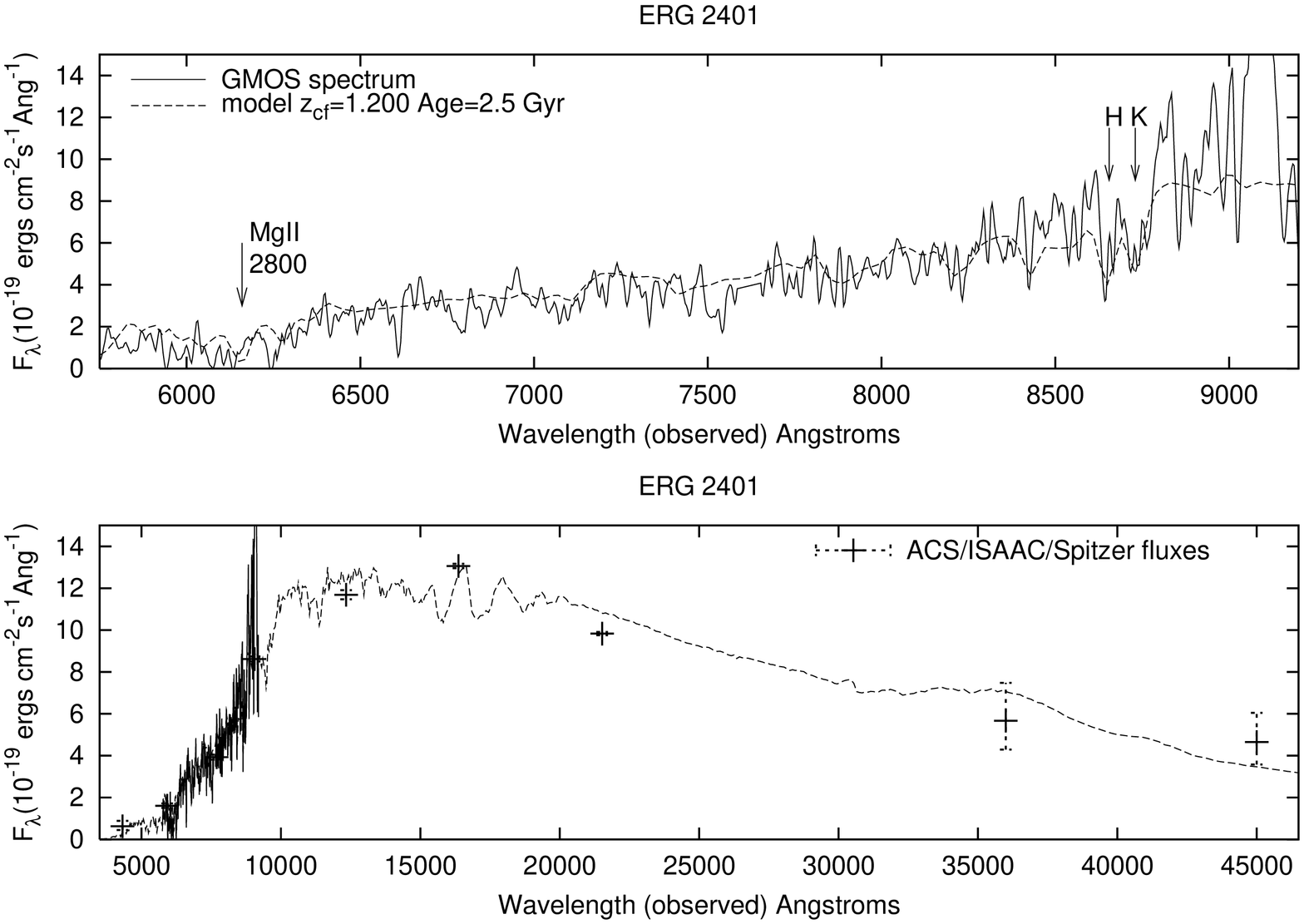,width=90mm,angle=-90}
\end{figure}
\begin{figure}
\psfig{file=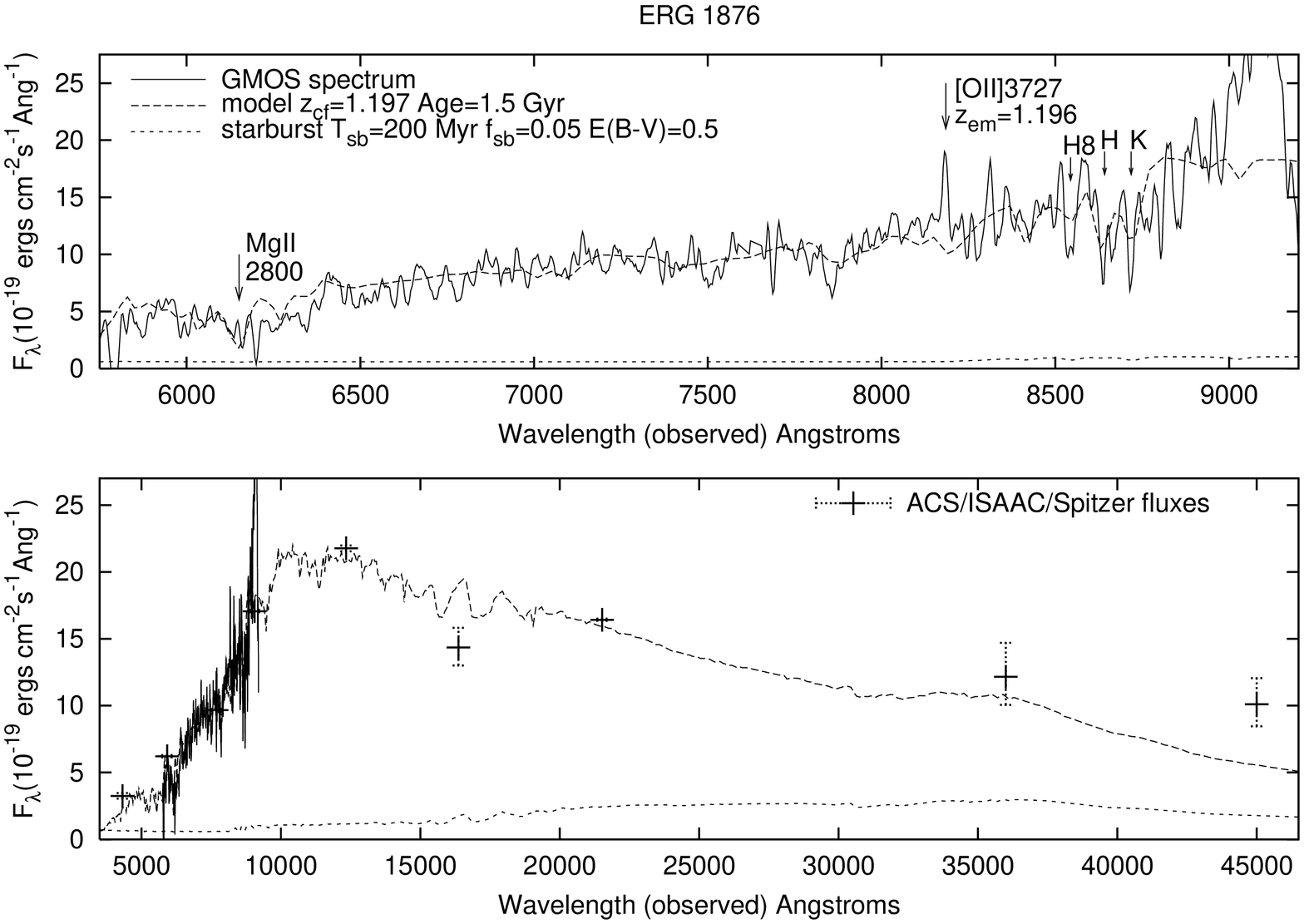,width=90mm,angle=-90}
\end{figure}
\begin{figure}
\psfig{file=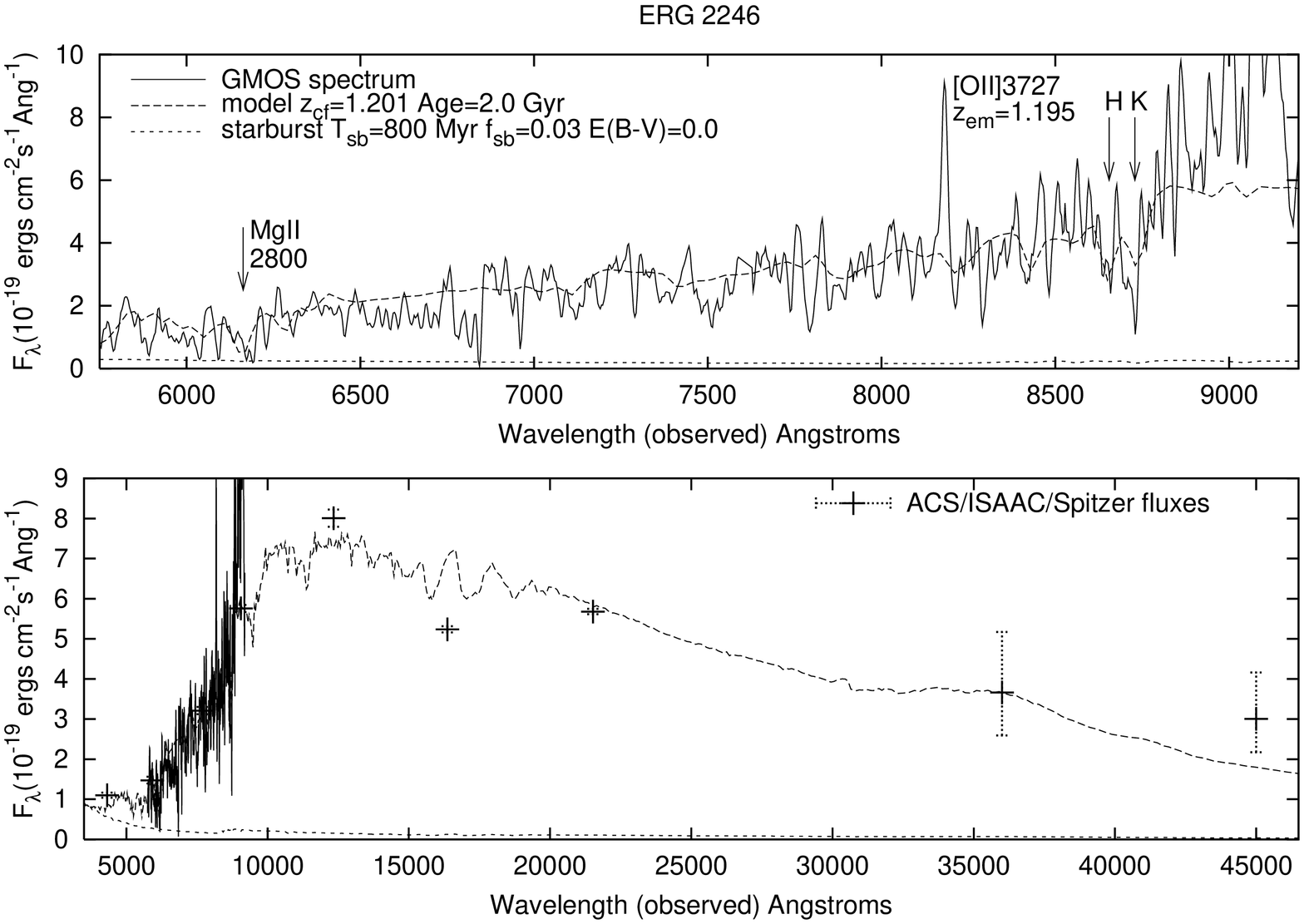,width=90mm,angle=-90}
\end{figure}
\begin{figure}
\psfig{file=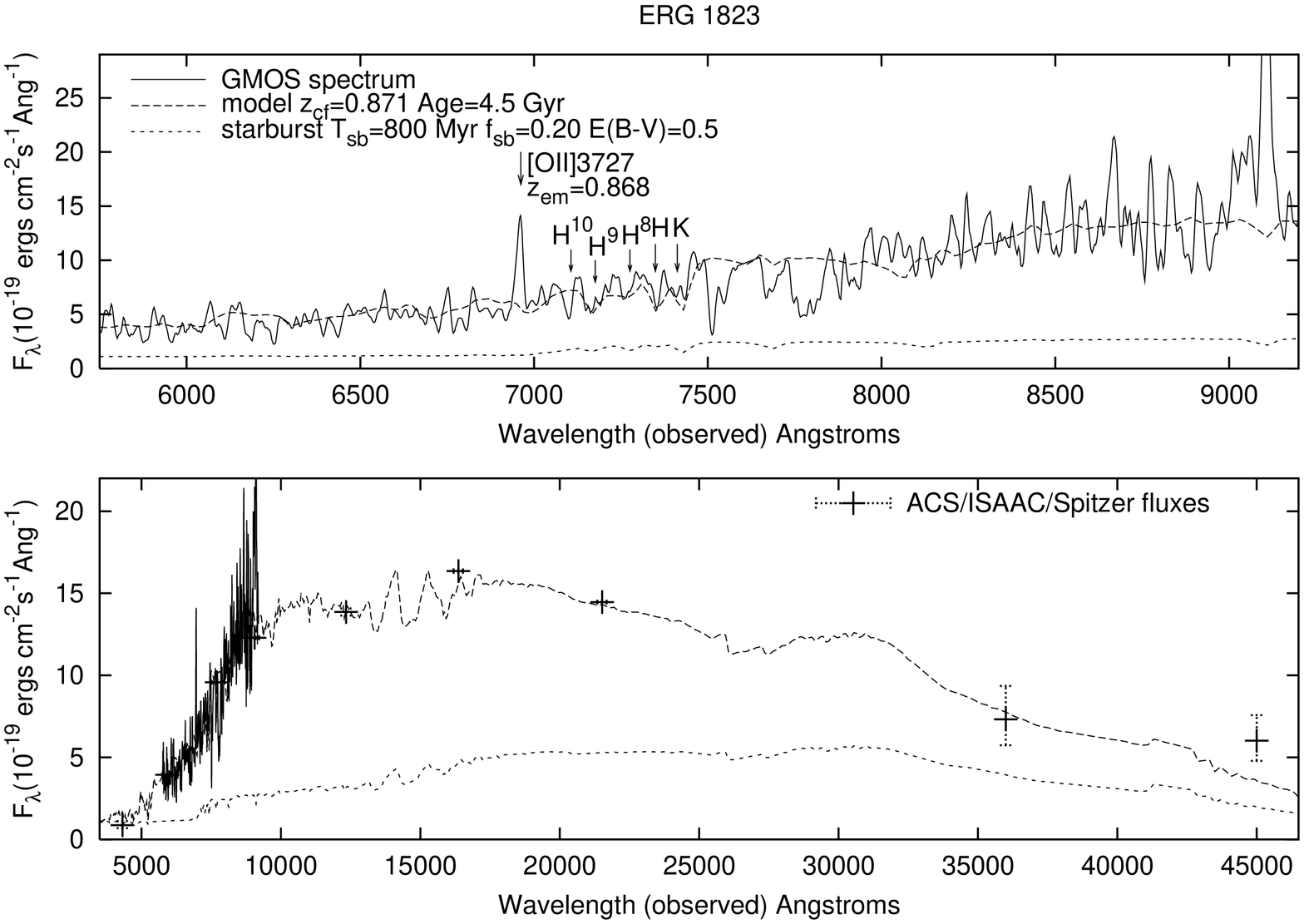,width=90mm,angle=-90}
\end{figure}
\begin{figure}
\psfig{file=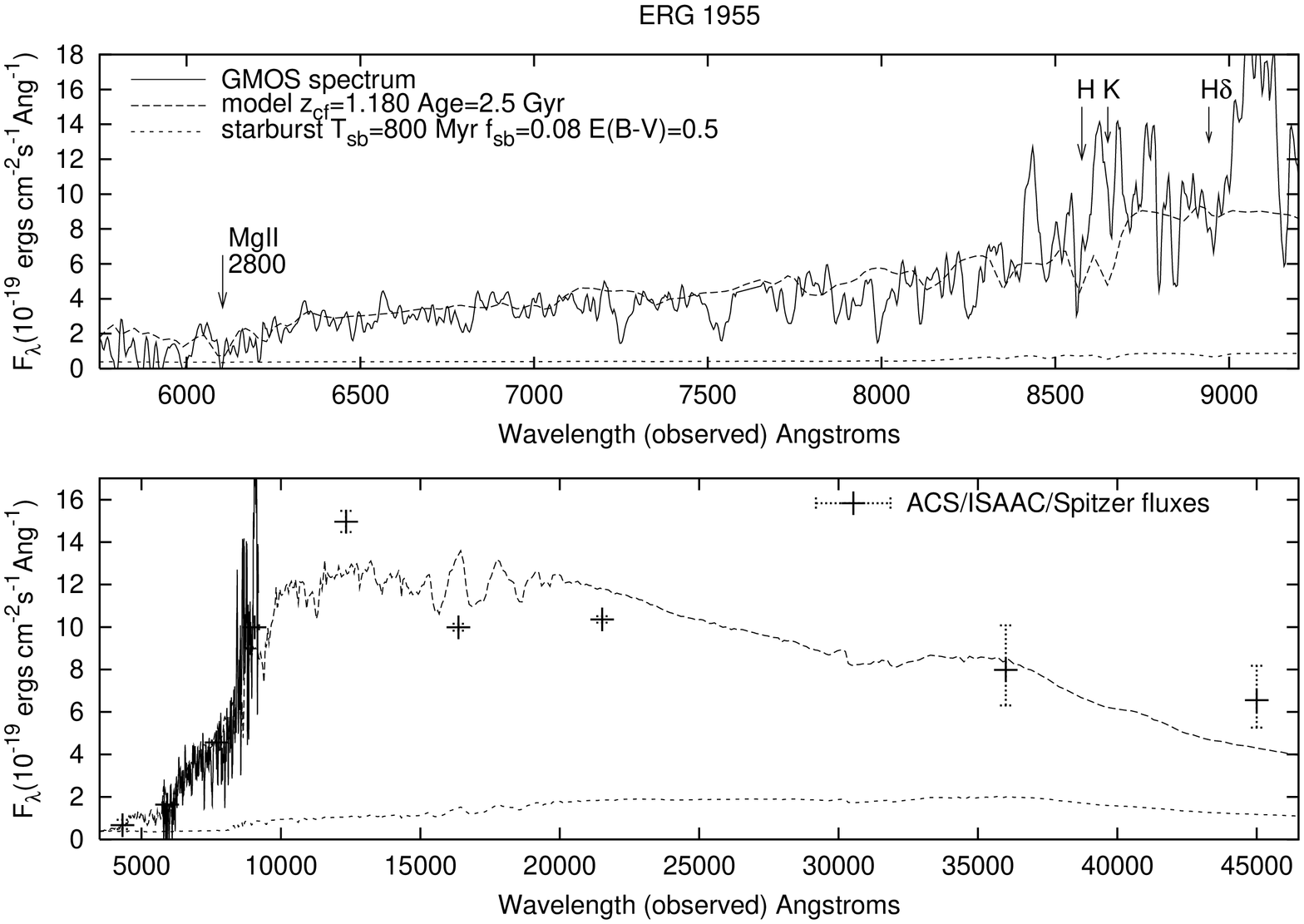,width=90mm,angle=-90}
\end{figure}
\begin{figure}
\psfig{file=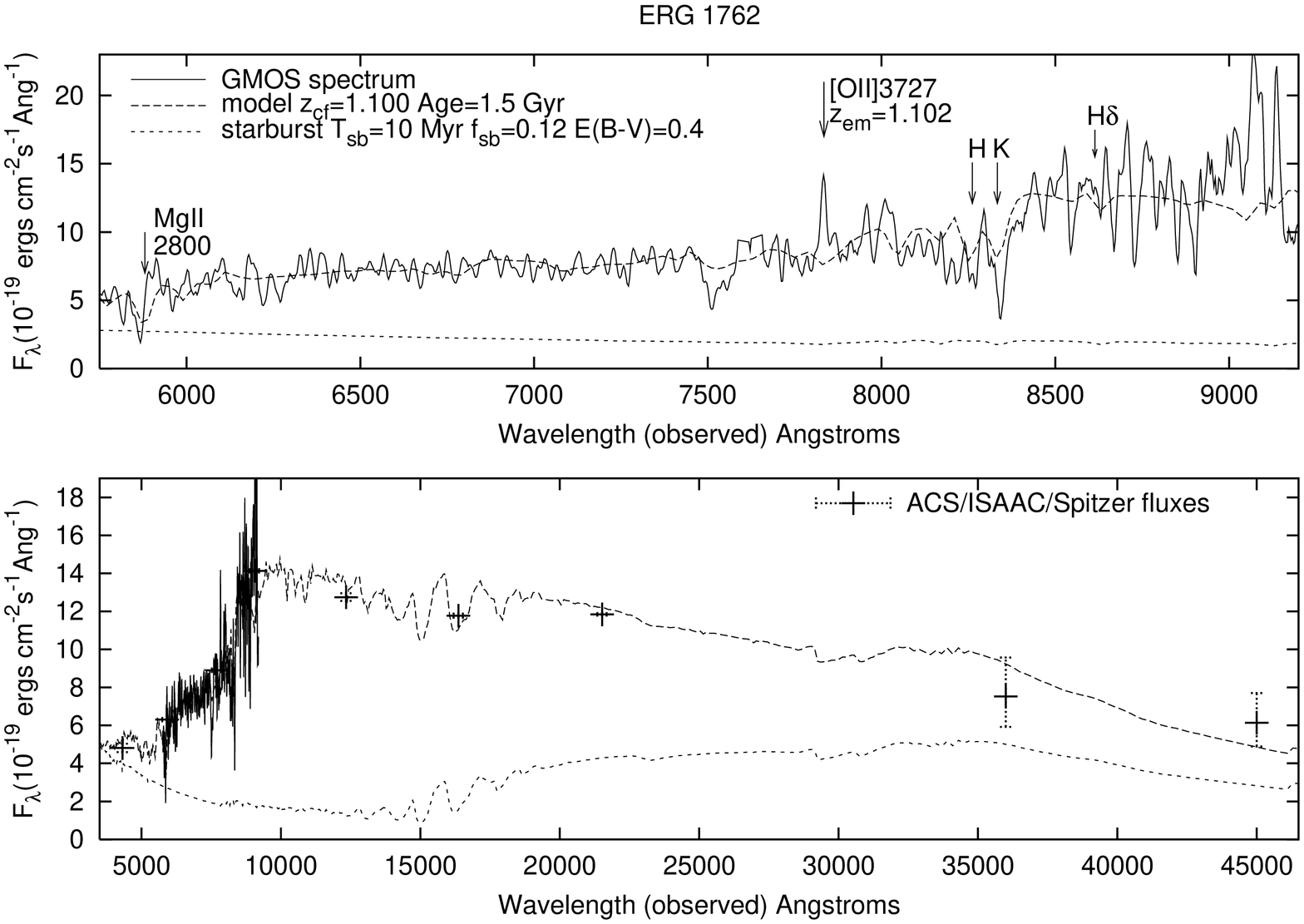,width=90mm,angle=-90}
\end{figure}
\begin{figure}
\psfig{file=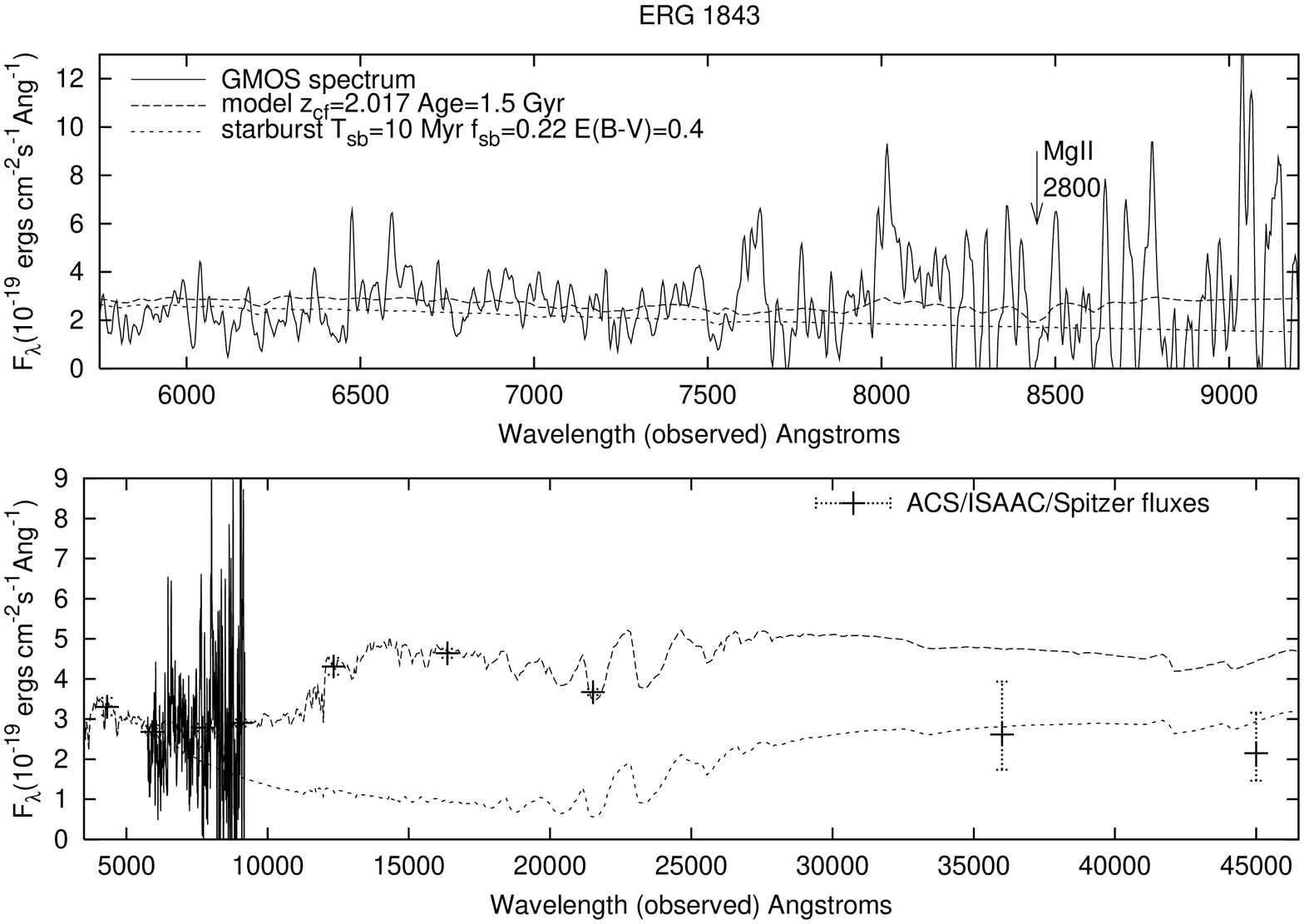,width=90mm,angle=-90}
\end{figure}
\begin{figure}
\psfig{file=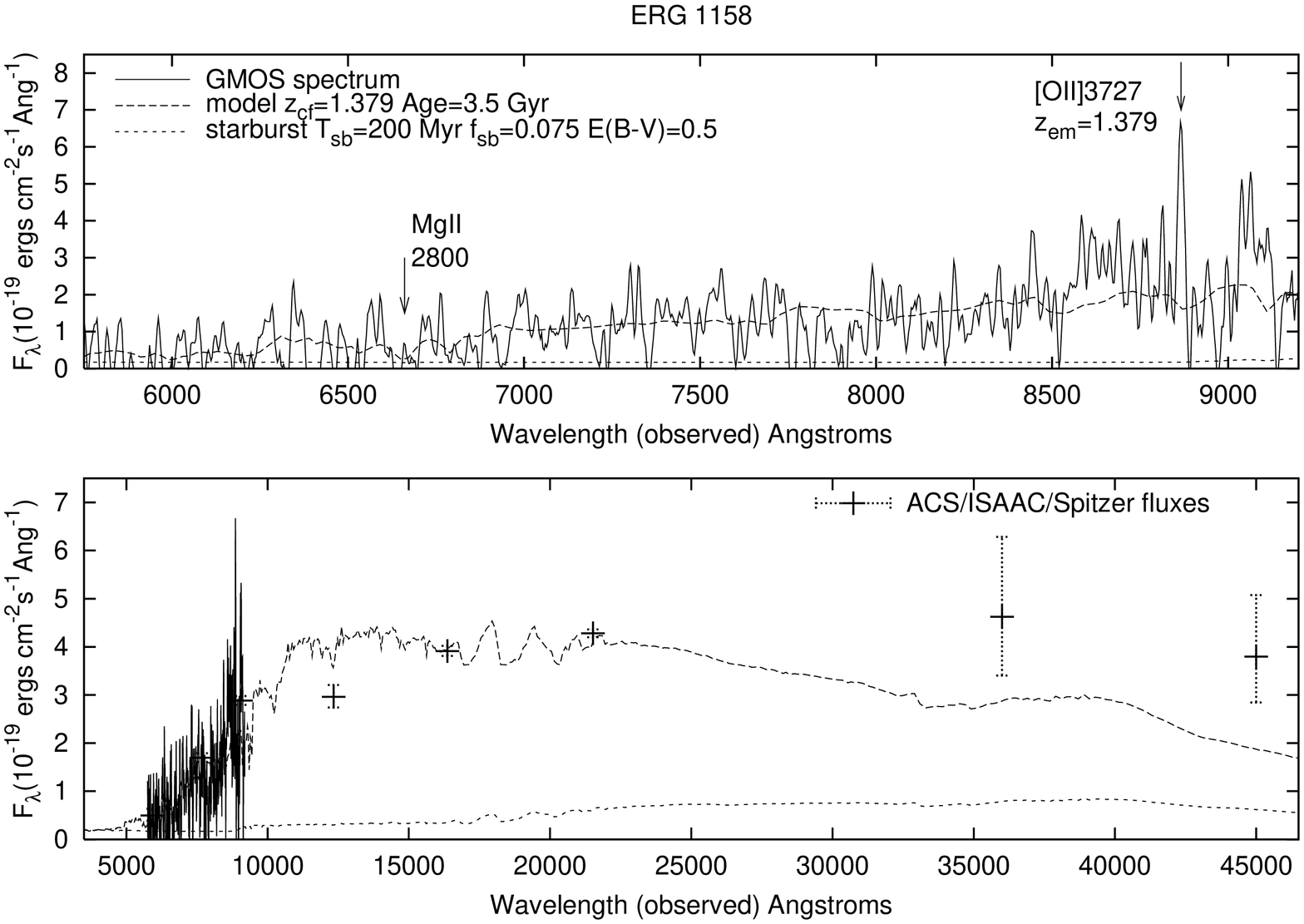,width=90mm,angle=-90}
\caption{GMOS spectra and broad-band (ACS/ISAAC/Spitzer)
 photometry  of 16 ERGs, calibrated in
$F_{\lambda}$. For each the upper plot shows the
GMOS spectrum (solid line), our  best-fit two-component
model (dotted), and, where non-zero, the starburst component of the model.
 Labelled are the age of the model passive component, and the timescale $T_{sb}$,
  fractional flux contribution at $\lambda_{rest}=4500\rm \AA$, $f_{sb}$ and dust
   reddening $E(B-V)$ of the starburst component, and obvious spectral features.
The lower plot shows a wider wavelength range with
the 9 broad-band fluxes,
the GMOS spectrum (solid) and the best-fit models as in the upper plot.
Both plots are normalized to the 2.0 arcsec aperture fluxes.}
\end{figure}

\newpage
\onecolumn
\begin{table}
\caption{Redshift $z_{spec}$ (from emission lines, $z_{em}$, where
detected, otherwise from continuum fits, $z_{cf}$); best-fit age
of the passively evolving component ($T_{pas}$);
 fraction $f_{sb}$ of the flux at $\lambda_{restframe}=\rm 4500\AA$
 produced by the starburst component; age $T_{sb}$ of the constant-SFR starburst; dust
 reddening $E(B-V)$ of the starburst (magnitudes); mean stellar
 formation redshift ($z_{msf}$) corresponding
to the sum of $T_{pas}$ and  the lookback time of $z_{spec}$; and the
 best-fit model mass
for the passive ($M_{pas}$) and starburst ($M_{sb}$) stellar
populations (normalized to the 2.0 arcsec aperture fluxes), in units of the solar mass
$\rm M_{\odot}$; $Mf_{sb}$, the
 starburst component as a fraction of the total stellar mass; $M_{B}$ the
 total absolute magnitude in rest-frame $B$ (AB system). Errors are
 approximately $90\%$ confidence intervals.}
\begin{tabular}{lccccccccc}
\hline
\smallskip
ID no. & $z_{spec}$ & $T_{pas}$ (Gyr) & $f_{sb}$ & $T_{sb}$ (Myr) & $E(B-V)$
& $z_{msf}$ & $M_{pas}$ & $M_{sb}$ & $Mf_{sb}$ \\
1152 & 1.310 & $1.5^{+0.1}_{-0.6}$ & $0.035_{-0.03}^{+0.01}$ & $800_{-780}^{+0}$ &
 $0.0^{+0.1}_{-0}$ &  $1.95^{+0.06}_{-0.30}$ & $1.58\times 10^{11}$ & $7.36\times 10^8$ &
  0.0046   \\
1029 & 1.221 & $1.0^{+0.5}_{-0.2}$ & $0.105_{-0.075}^{+0.03}$ & $600^{+200}_{-560}$ &
$0.0^{+0.05}_{-0}$ & $1.68^{+0.25}_{-0.09}$ & $3.23\times 10^{10}$ & $6.58\times 10^8$ &
0.0199  \\
1412 & 1.092 & $1.0^{+0.2}_{-0.1}$ & $0.06_{-0.03}^{+0.045}$ & $800_{-200}^{+0}$ &
$0.5^{+0}_{-0.1}$ & $1.39^{+0.08}_{-0.04}$ & $4.70\times 10^{10}$ & $6.22\times 10^{9}$ &
0.1168 \\
1777 & 1.223 & $2.0\pm0.2$ & $0.135_{-0.045}^{+0.005}$ & $800_{-400}^{+0}$ &
$0.5^{+0}_{-0.05}$ & $2.08^{+0.13}_{-0.12}$ & $1.08\times 10^{11}$ & $1.70\times 10^{10}$ &
0.1355  \\
2536 & 1.512 & $0.6\pm 0.05$ & $0.04^{+0.03}_{-0.01}$ & $10^{+30}_{-0}$ & $0.5^{+0}_{-0.05}$
& $1.78\pm0.03$ & $6.17\times 10^{10}$ & $8.20\times 10^8$ & 0.0131  \\
1064 & 1.548 & $1.0\pm 0.15$ & $0.09^{+0.005}_{-0.03}$ & $40^{+60}_{-30}$ & $0.2\pm 0.1$ &
$2.04^{+0.10}_{-0.09}$ & $1.47\times 10^{11}$ & $1.31\times 10^9$ & 0.0089  \\
1143 & 1.254 & $4.5^{+1.1}_{-2.1}$ & $0.129_{-0.042}^{+0.014}$ & $10_{-0}^{+10}$ &
 $0.3_{-0.1}^{+0.05}$ & $7.23^{+\infty}_{-4.79}$ & $7.29\times 10^{10}$ & $1.89\times 10^8$ &
0.0026 \\
1565 & 1.108 & $2.0^{+0.2}_{-0.6}$ & $0.090_{-0.015}^{+0.045}$ & $10_{-0}^{+200}$ &
$0.5^{+0}_{-0.05}$ & $1.85^{+0.10}_{-0.28}$ & $3.87\times 10^{10}$ & $3.38\times 10^8$ &
 0.0087 \\
2401 & 1.200 & $2.5^{+0.2}_{-0.1}$ & $0.0_{-0}^{+0.015}$ & - & - & $2.37^{+0.17}_{-0.07}$ &
$1.63\times 10^{11}$ & 0 & 0 \\
1876 & 1.196 & $1.5^{+0.3}_{-0.1}$ & $0.05^{+0.04}_{-0.02}$ & $200^{+600}_{-140}$ &
 $0.5_{-0.1}^{+0}$ &  $1.75^{+0.16}_{-0.04}$ & $1.94\times 10^{11}$ & $5.10\times 10^9$
  & 0.0256  \\
2246 & 1.195 & $2.0^{+0.5}_{_0.2}$ & $0.03\pm0.024$ & $800_{-790}^{+0}$ &
$0.0^{+0.2}_{-0.0}$ & $2.02^{+0.34}_{-0.11}$ & $8.27\times 10^{10}$ &
 $2.64\times 10^8$ & 0.0032\\
1823 & 0.868 & $4.5^{+0.7}_{-1.7}$ & $0.20^{+0.025}_{-0.05}$ & $800^{+0}_{-100}$ &
$0.5^{+0}_{-0.05}$ & $2.98^{+1.09}_{-1.25}$ & $1.91\times 10^{11}$ &  $2.36\times 10^{10}$
& 0.1100  \\
1955 & 1.180 & $2.5^{+0.2}_{-0.3}$ & $0.08_{-0.05}^{+0.01}$ & $800^{+0}_{-200}$ &
$0.5^{+0}_{-0.1}$ & $2.32^{+0.16}_{-0.21}$ & $1.43\times 10^{11}$ & $1.07\times 10^{10}$ &
0.0695 \\
1762 & 1.102 & $1.5_{-0.5}^{+0.1}$ & $0.12^{+0.015}_{-0.005}$ & $10^{+5}_{-0}$ & $0.4\pm0.05$
& $1.60^{+0.04}_{-0.19}$ & $9.35\times 10^{10}$ & $8.89\times 10^8$ & 0.0094 \\
1843 & 2.017 & $1.5_{-0.7}^{+0.1}$ & $0.22^{+0.02}_{-0.04}$ & $10^{+5}_{-0}$  & $0.4\pm 0.05$ &
$3.42^{+0.16}_{-0.81}$ & $1.85\times 10^{11}$ & $3.63\times 10{9}$ & 0.0193 \\
1158 & 1.379 & $3.5_{-1.5}^{+0.6}$ & $0.075^{+0.045}_{-0.060}$ & $200^{+600}_{-180}$ &
 $0.5^{+0}_{-0.1}$ &
$4.59^{+2.53}_{-2.17}$ & $1.21\times 10^{11}$ & $2.22\times 10^9$ & 0.0183\\
\hline
\end{tabular}

$^*$ Host galaxy of obscured AGN --
age/mass estimates may be affected by non-stellar component.
\end{table}

\begin{table}
\caption{Nine-band (HST-ACS, {\it ISAAC} and Spitzer IRAC)
photometry of the 16 ERGs in the redshift sample,
 with fluxes measured in fixed 2.0 arcsec diameter circular
 apertures (2.8 arcsec for Spitzer data) and expressed in the AB magnitude system,
 where $m_{AB}=-48.60-2.5$ log
$F_{\nu}$ and $1\mu\rm Jy$ is $m_{AB}=23.90$. Absolute magnitude
in rest-frame $B$ (AB system), as estimated from the model fit to the aperture
 photometry ($M_B$), and with an approximate correction to total magnitude
 $M_{BT}$.}
\begin{tabular}{lccccccccccc}
\hline
\smallskip
ID no. & $B_{435}(AB)$ & $V_{606}(AB)$ & $I_{775}(AB)$ & $Z_{850}(AB)$ &
$J(AB)$ & $H(AB)$ & $K_s(AB)$ & $3.6\mu\rm m$ & $4.5\mu \rm m$ & $M_B$ & $M_{BT}$\\
1152 & 26.06 &  25.13 &  23.83 &  22.88 &  21.83 &  21.43 & 21.16 & 20.21  & 19.93
 &   -21.51 & -21.69 \\
1029 & 25.85 &  25.56 &  24.58 &  23.62 &  22.79 &  22.32 & 22.11  & 21.27 &  21.00
& -20.43 & -20.49 \\
1412 & 26.03 &  24.94 &  23.70 &  22.84 &  22.06 &  21.65 & 21.21 & 20.29  & 20.02
& -20.77 & -20.91 \\
1777 & $(n.d).$ & 25.13 &  24.07 &  23.11 &  22.06 &  21.75 & 20.90 & 20.04 & 19.79
& -20.95 & -21.14 \\
2536 &  25.71 &  25.04 & 24.07 &  23.47 &  22.48 &  22.10 &  21.51 & 20.79 & 20.52
& -21.70 & -21.77 \\
1064 & 25.14 &  24.76 &  23.92 &  23.12 &  21.78 &  21.57 & 20.94 & 20.08  & 19.81
& -22.06 & -22.18\\
1143 & 27.28 &  25.47 &  25.40 &  24.78 &  23.33 &  22.76 & 22.15 & 21.37  & 21.10
  & -19.73 & -19.81 \\
1565 & 27.08 &  25.92 &  24.50 &  23.73 &  23.11 &  22.56 & 21.69 & 20.88  & 20.61
& -19.79 & -19.87 \\
2401 & 27.42 &  25.71 &  24.18 &  22.97 &  21.97 &  21.23 & 20.95 &  20.43 & 20.16
 & -21.03  & -21.21\\
1876 & 25.64 &  24.25 &  23.20 &  22.23 &  21.29 &  21.13 & 20.39 & 19.60  & 19.31
& -21.75 & -22.06\\
2246 & 26.81 &  25.81 &  24.39 &  23.41 &  22.38 &  22.22 & 21.54 & 20.90  & 20.63
 & -20.53 & -20.55 \\
1823 & 27.07 &  24.74 &  23.21 &  22.58 &  21.78 &  20.99 & 20.53 & 20.15 & 19.88
& -20.86 & -20.94\\
1955 & 27.35 &  25.69 &  24.01 &  22.81 &  21.70 &  21.52 & 20.89 & 20.06 & 19.78
& -20.99 & -21.13 \\
1762 & 25.21 &  24.23 &  23.29 &  22.43 &  21.87 &  21.34 & 20.74 & 20.12 &  19.86
& -21.06 & -21.06 \\
1843 & 25.62 &  25.16 &  24.55 &  24.15 &  23.05 &  22.36 & 22.02 & 21.27 & 21.00
& -21.96 & -22.14 \\
1158 & $(n.d).$ &  26.99 &  25.09 &  24.16 &  23.46 &  22.54 & 21.85 & 20.65 & 20.38
& -20.35 & -20.58 \\
\hline
\end{tabular}

(n.d)$=$ not detected (too faint) in this passband.
\end{table}
\onecolumn
\twocolumn

\begin{table}
\caption{Equivalent width of the [OII] $3727\rm \AA$ emission line, converted to restframe,
the line flux ($f_{\rm OII}$) and
luminosity ($L_{\rm OII}$), and the star-formation rate derived from this
($\rm SFR_{OII}=1.4\times 10^{-41}L_{\rm OII}$ $\rm M_{\odot}yr^{-1}$), for
ERGs in our redshift sample (these quantities
normalized to the  ERG flux in a 2.0 arcsec diameter aperture). $1\sigma$ upper
limits are given for galaxies where no [OII] emission is seen.}
\begin{tabular}{lcccc}
\hline
ID no. & $\rm EW[OII]$ & $f_{\rm OII}$ & $L_{\rm OII}$ & $\rm SFR_{OII}$ \\
   & $\rm \AA$  & $10^{-18}$ ergs & $10^{40}$  &
$\rm M_{\odot}yr^{-1}$ \\
 & restframe  & $\rm s^{-1} cm^{-2}$ & ergs $\rm s^{-1}$ &  \\
\smallskip
1152 & 5 &  $8.40\pm 2.90$ & $9.21\pm 3.18$ & $1.29\pm 0.45$ \\
1029 & 46 &  $38.1\pm 2.8$ & $35.1\pm 2.5$ & $4.91\pm 0.35$ \\
1412 & 9 & $11.34\pm 2.45$ & $7.88\pm 1.70$ & $1.10\pm 0.22$ \\
1777 &   & $<3.33$ & $<3.08$ & $<0.43$  \\
2536 & ?  & ? & ? & ? \\
1064 & ?  & ? & ? & ? \\
1143 & 21 &  $7.38\pm 2.13$ & $7.21\pm 2.08$ & $1.01\pm0.29$ \\
1565 & 31 &  $15.85\pm 2.34$ & $11.43\pm 1.69$ & $1.60\pm 0.24$ \\
2401 &   & $<2.41$ & $<2.12$ & $<0.30$  \\
1876 & 5 & $10.26\pm 2.76$ & $8.94\pm 2.35$ & $1.25\pm 0.33$ \\
2246 & 12 & $10.22\pm 1.57$ & $8.91\pm 1.37$ & $1.25\pm 0.19$ \\
1823 & 10 & $18.58\pm 1.65$ & $7.26\pm 0.65$ & $1.02\pm 0.08$ \\
1955 &    &  $<2.20$ & $<1.86$ & $<0.26$ \\
1762 & 8 &  $10.48\pm 3.32$ &   $7.45\pm 2.35$ & $1.04\pm 0.33$ \\
1843 & ?  & ? & ? & ? \\
1158 &  40 &  $10.51\pm 3.36$ &   $13.13\pm 4.20$ & $1.84\pm 0.59$ \\
\hline
\end{tabular}

?: Galaxy is at $z>1.5$ where $\rm [OII]3727\AA$ would not be visible
in GMOS $\lambda$ range.

\end{table}
\section{Morphology and Size of ERGs in the Redshift Sample}
The HST-ACS images, in the stacked
form used here (version V1.0 from the STScI),
 have  a pixelsize 0.03 arcsec and resolution
$\rm FWHM\simeq 0.08$ arcsec $(\simeq 0.66$ kpc at $z=1.2$).
Figure 5 shows all 16 galaxies as they appear on  the $I_{775}$-band  ACS
image, and Figure 6 the $B_{435}$-band images
 ($\lambda_{restframe}\sim 2000\rm \AA$), which have poorer
signal-to-noise but provide extra information by highlighting
 regions of
star-formation (SF).

The appearance of each ERG is described briefly below
(by `high-SB' we mean an $I_{775}$-band central
surface brightness of
about 0.033 counts $\rm pixel^{-1} s^{-1}=21.7$ AB mag $\rm
arcsec^{-2}$, a value typical
 of the spheroidal ERGs).

1152 ([OII] emitter)
is a high-SB spheroidal surrounded by a much fainter ring (post-merger?). Only
the central nucleus is
 visible in the $B$ image.

1029 ([OII] emitter) appears in $I$ to be a normal spheroidal, with a
bulge profile.  However,
the $B$ image shows spots within the galaxy which
could be SF regions.

1412 ([OII] emitter) is a large ($\sim 1.8$ arcsec diameter) disk with spiral arms
 and      a small central nucleus. In $B$ the nucleus cannot be seen but
 two brighter regions in the outer arms are just visible.

1777 (no [OII]) is a barred spiral of large diameter ($\sim 2.4$ arcsec) and
 relatively low SB. A
nucleus and two spiral arms are clearly visible in $I$. It does not appear
 disturbed. In the $B$ image the nucleus and arms cannot be seen and all that
 is visible is a tiny bright spot within the N spiral arm.

2536 ($z>1.5$; UV-bright) is a merging pair of separation
$\sim 0.25$ arcsec. The NW component is  smaller but
higher in SB,
the SE is irregular with bright regions on the W and SE sides.
In the $B$ band both galaxies are visible, but only the W and SE parts of the
irregular component and the centre of the NW galaxy are prominent. Both
 galaxies appear to contain bright SF regions within redder
 disks/spheroids. This ERG has
 the youngest $T_{pas}$ of our sample combined with a young ongoing starburst.

1064 ($z>1.5$; UV-bright) appears in the $I$ band to be a high-SB elliptical.
       However, in  the $B$-band
 the nucleus disappears and is replaced by bright regions
 on either side of it, forming a bar. This bar is very  asymmetric and there
       is no sign of spiral arms,
 implying that this galaxy
  is not a barred spiral but a  merger.

1143 ([OII] emitter) is a very small speroidal, without visible features (in
both $I$ and $B$) but appears to be a dusty starburst.

1565 ([OII] emitter) is a complex system, with four separate intensity peaks.
  The southernmost galaxy has the highest SB in
$I$ and could be a spheroid. Only one of the 4 components,
an amorphous galaxy at the NW, is visible in $B$. This could be a
multiple merger of a star-forming and 3 passive galaxies.

2401 (no [OII]) is a small, roundish, high-SB spheroid. It is not visible in
the $B$-band. The spectrum gives no evidence of ongoing or recent SF.

1876 ([OII] emitter) appears to be a large, face-on spiral
with a high-SB nucleus,
  and two arms reaching to a radius $\simeq 0.9$ arcsec.
The galaxy is ringed by bright regions, presumably
star-forming hotspots on the basis of their visibility in the
 $B$ image, with gaps in the ring at the N and S.
This galaxy has a strong colour gradient; the nucleus ($r<0.5$ arcsec)
has a `passive'
colour of
$(B-I)_{AB}=2.49$, but the ring ($0.5<r<1.2$ arcsec) has  $(B-I)_{AB}=1.39$,
 a colour similar to the bluest
 galaxies in this sample (1029, 1064 and the AGN 1843) and indicating an
observer-frame visible flux
dominated by young stars.
 This very interesting object
is a red spiral with ongoing SF, apparently of long duration,
 concentrated in the
outermost rim.

2246 ([OII] emitter) is a small, roundish, high-SB spheroid. In the $B$ band
it looks even more smaller, and asymmetric, suggesting a central SF region.
The SF appears to be prolonged and this is probably a post-merger.

1823 ([OII] emitter) is a merger of nuclear separation  $\simeq
0.3$ arcsec,  apparently of a spheroid (N)
and a more
irregular galaxy (S)  with a lower $I$-band SB and three intensity peaks.
 In $B$
the N spheroid is barely visible (passive?) but the S irregular is bright
(starburst?).

1955 (no [OII]) is a  high-SB spheroid with slight asymmetry in the outer
regions, possibly due to perturbation by
a smaller neighbouring galaxy separated by  0.8 arcsec separation.
 $B$ morphology is similar but faint.

 1762  ([OII] emitter)  resembles several of the disk mergers in the Arp catalog.
The $I$ image shows two nuclei
 separated by 0.34 arcsec, of which the larger (N)
is at the centre of a
partial ring or curved tail of
 radius $\simeq 0.4$ arcsec, and the smaller (S) has a long ($\leq 1.0$ arcsec)
straight tail.
In the $B$ band the system appears much more fragmented, the
large nucleus is less prominent but
the S nucleus is bright and parts of the tails are visible, together with a
 bright region in the  SW. This is a merger
 with extensive SF, and some of the bright
 clumps in the tails might be new dwarf galaxies being formed in
 the interaction.
\onecolumn
\begin{figure}
\psfig{file=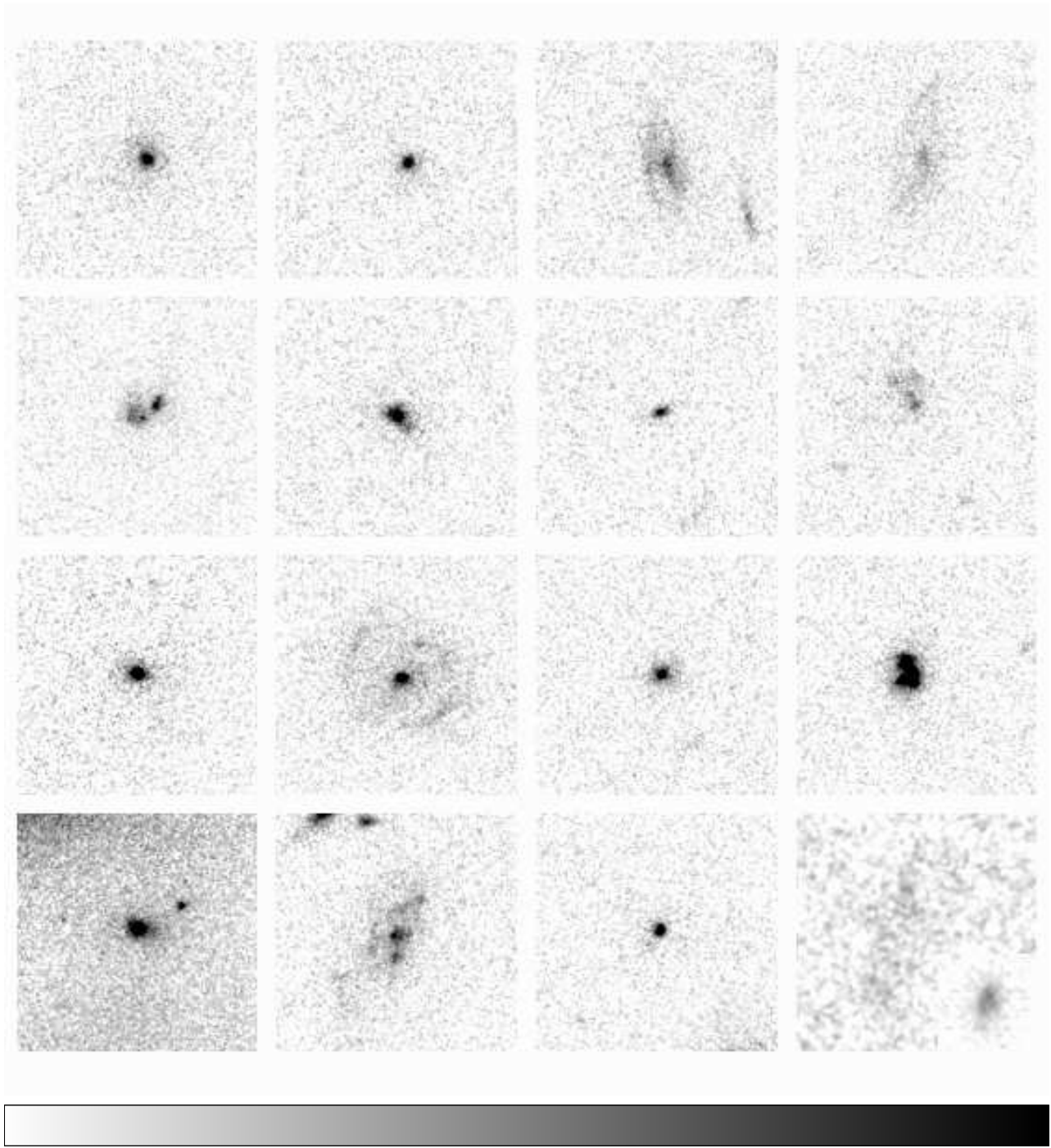,width=170mm}
\caption{ACS $I_{775}$-band images, from the GOODS data version v1.0,
 of the 16 ERGs in our redshift sample;
(top row, left to right) 1152, 1029, 1412, 1777, (2nd row) 2536, 1064, 1143,
 1565, (3rd row) 2401, 1876, 2246, 1823, (bottom row) 1955, 1762, 1843, 1158.
Each picture shows a $3.9\times 3.9$ arcsec area, with an inverse greyscale
 with black corresponding to 22.27 AB mag $\rm arcsec^{-2}$. Galaxy 1158,
due to its very low SB, is shown smoothed by a $\rm FWHM=0.07$ arcsec Gaussian
and scaled
 up in  intensity by a factor 2, and its ISAAC
$K_{s}$-band image is shown inset in the lower right (at a scale 0.4 that of
 its ACS image).}
\end{figure}
\begin{figure}
\psfig{file=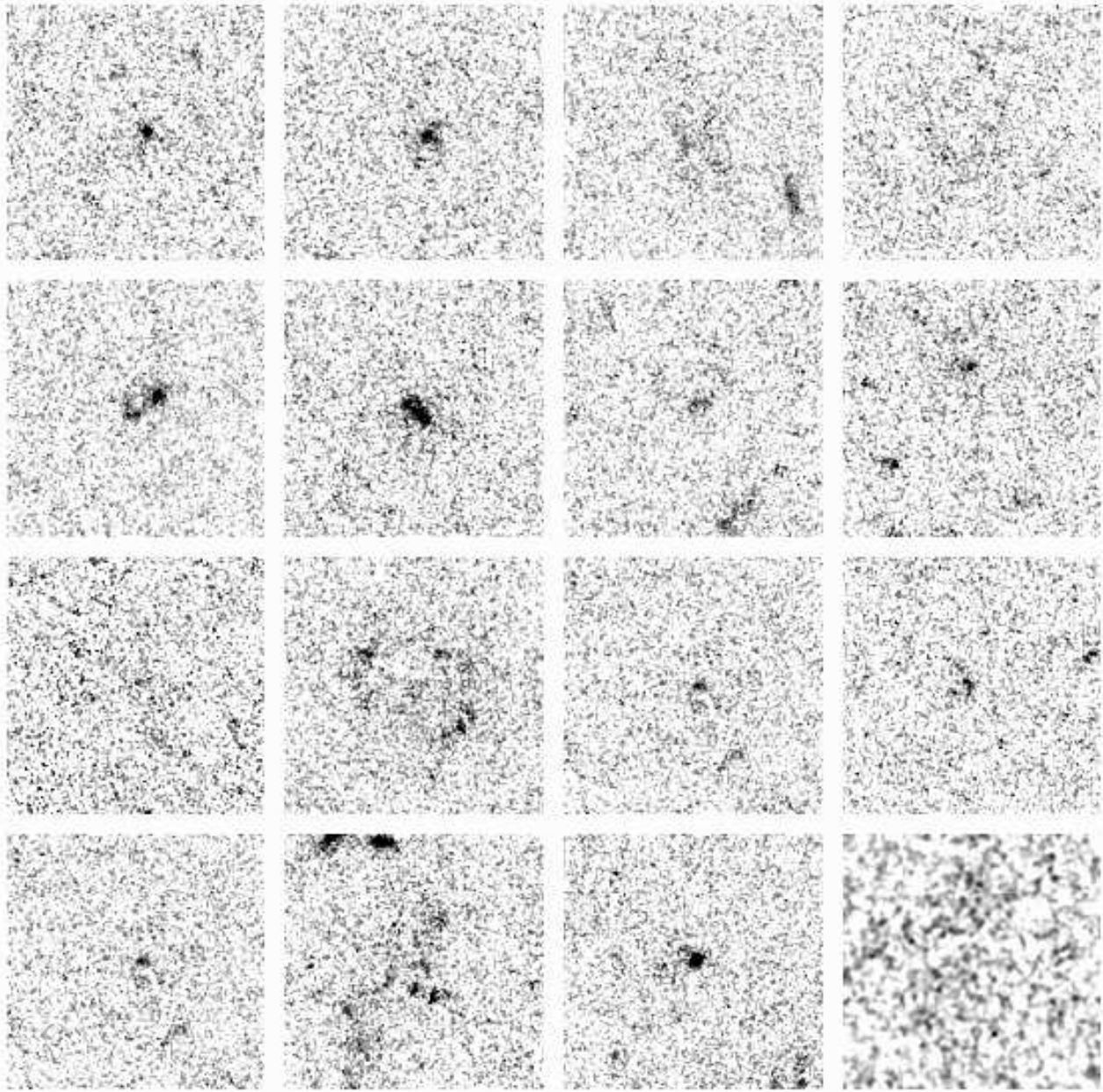,width=170mm}
\caption{ACS $B_{450}$-band images, from the GOODS data version v1.0,
 of the 16 ERGs in our redshift sample;
(top row, left to right) 1152, 1029, 1412, 1777, (2nd row) 2536, 1064, 1143,
 1565, (3rd row) 2401, 1876, 2246, 1823, (bottom row) 1955, 1762, 1843, 1158.
Each picture shows a $3.9\times 3.9$ arcsec area,  with an inverse greyscale
 with black corresponding to 24.03 AB mag $\rm arcsec^{-2}$. Galaxy 1158,
due to its very low SB,
 is shown smoothed by a $\rm FWHM=0.07$ arcsec Gaussian and scaled
 up in  intensity by a factor 2.}
\end{figure}
\twocolumn
1843 ($>1.5$, {\it Chandra} source, AGN) has a central SB
  twice  that of the other spheroidals in this sample. It appears to be a
  bright  point-source within a faint extended host galaxy
  visible in both $I$ and $B$.

1158 ([OII] emitter) is of very low SB in the $I$-band. It is more prominent
on the $K$-band  image.
 It appears to be an almost edge-on disk or ring some 1.6
arcsec in long-axis  diameter, without a bright nucleus and without much
 asymmetry. In $B$ it is even fainter and almost invisible, even with
 smoothing, but probably of similar size and shape.
 It is most likely a red spiral
galaxy with the `ring' appearence suggesting that
most SF is in the outer regions rather than the nucleus (as in 1876?).
\begin{table}
\caption{Half-light radius and Sersic index of the ERGs from model
profiles  fitted to HST-ACS $Z_{850}$-band images. Galaxy 1876 is fitted as a
combination disk ($n=1$) + bulge ($n=4$), and galaxy/AGN 1843 as a combination
bulge + point-source ($n=0$).}
\begin{tabular}{lccc}
\hline
\smallskip
ID no. & Sersic Index $n$ & $r_{hl}$ (arcsec) & $r_{hl}$ (kpc) \\
1152 & 4.8 & 0.969 & 8.43    \\
1029 & 5.1 & 0.330 & 2.84    \\
1412 & 1.2 & 0.402 & 3.39    \\
1777 & 1.0 & 0.543 & 4.68    \\
2536 & 1.4 & 0.321 & 2.83    \\
1064 & 3.5 & 0.360 & 3.18    \\
1143 & 2.4 & 0.054 & 0.47    \\
1565 & 1.4 & 0.384 & 3.25    \\
2401 & 5.8 & 0.411 &  3.53    \\
1876 & (85\%) 1.0 & 0.969 &  8.32    \\
     & (15\%) 4.0 & 0.060 & 0.52    \\
2246 & 4.2 & 0.261 & 2.24      \\
1823 & 1.2 & 0.258 & 2.05     \\
1955 & 4.4 & 0.807 & 6.92      \\
1762 & 1.4 & 0.582 & 4.92      \\
1843 & (60\%)  4.0 & 0.675 &  5.91 \\
     & (40\%)  0.0 & - & -  \\
1158 & 1.0 & 0.864 & 7.56  \\
\hline
\end{tabular}
\end{table}
\medskip
Sizes (half-light radii, $r_{hl}$) were estimated
for the ERGs by fitting the profile of each
in the HST-ACS $Z_{850}$-band (approximately the rest-frame $B$-band) with
the Sersic profile  $I(r)=I_o=exp[({r\over r_0})^{1\over n}]$, using a method
in which this was convolved with
 the detailed form of the HST-ACS point-spread function
(see Floyd et al. 2004). In this equation $n$ is the Sersic index, $n=1$ for
an exponential (pure disk) profile and $n=4$ for a de Vaucouleurs bulge
profile. The half-light radius $r_{hl}$ is a function of $r_0$ and $n$.
Table 4 gives PSF-corrected $r_{hl}$ and $n$ for the 16 ERGs; note that two
galaxies were not well-represented by a single Sersic profile and have been
fitted with two component models.

Morphologically, these ERGs can be placed into three broad
 classes:
 (i) spheroidals of high Sersic index $\langle n \rangle=4.39\pm 0.40$ (1152, 1029
1143, 2401, 2246, 1955, 1843),
(ii) red disk or spiral galaxies, of low Sersic index (exponential profiles)  $\langle n \rangle=1.05\pm 0.05$
and  without much sign of disturbance
(1412, 1777, 1876, 1158), and
 (iii)
merging or highly disturbed galaxies (2536, 1064, 1565, 1823, 1762)
typically with intermediate Sersic indices $\langle n \rangle=1.78\pm 0.43$.

Figure 6 shows the Sersic model $r_{hl}$ for the 16 ERGs plotted against the
estimated total absolute magnitude $M_B$ (the $M_{BT}$ from Table 3). Also plotted for comparison
are, firstly, a mean
$r_{hl}-M_B$ relation for local galaxies, as derived from the Driver et
al. (2005)  relation of $B$-band effective surface brightness to absolute
magnitude for
the Millenium Galaxy Catalog, converted to $H_0=70$ and AB magnitudes
 ($B_{AB}=B_{MGC}-0.13$). This sample is composed predominantly  of spirals.
Secondly, to represent local spheroidals we plot the best-fit size-luminosity
relation of early-type (Sersic $n>2.5$) galaxies in the Sloan Digital Sky
Survey (Shen et al. 2003),
with an approximate conversion $B_{AB}=r +1.17$ (Driver et al. 2005).

Amongst spheroidal and disk ERGs, there  is considerable scatter in the
size-luminosity relation, but generally little, or moderate ($<1$ mag) enhancement
 in SB relative to local
 galaxies of the same type. This would be
consistent with the passive evolution of ellipticals which formed Gyr earlier, or
with the exponetially declining SFRs of early-type spirals.
One ERG, the disk 1158, has a significantly lower SB than
a typical local galaxy and on the basis of its spectrum (Figure 3) this is
likely due to heavy internal dust extinction, as well as a diffuse
morphology.
Merging ERGs, in contrast, show in their size-luminosity relation a marked
(1--2 mag) SB brightening relative to local galaxies, especially in the case of
the  $z>1.5$ mergers 2536 and 1064,  presumably due to
 intense, short-term  starbursts.

\begin{figure}
\psfig{file=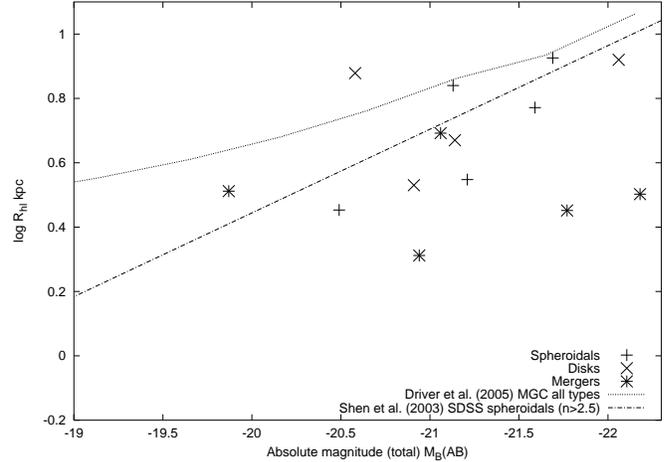,width=90mm,angle=-90}
\caption{Half-light radius from fitted Sersic model plotted against total absolute magnitude in $B_{AB}$,
 for the 16 ERGs in the redshift sample (symbols indicate
morphology). For the AGN 1843 only the 60\% extended component of the
flux is included in the plotted absolute magnitude (giving $M_B=-21.59$). Also plotted are
the mean size-luminosity relations  for galaxies in the Millenium
Galaxy Catalog, derived from Driver et al. (2005), and for early-type galaxies in the
Sloan Digital Sky Survey (Shen et al. 2003).}
\end{figure}

\section{ERG Clustering vs. Photometric Redshift}
We rexamine the clustering of ERGs in the CDFS sample using the photometric
redshift estimates  of Caputi et al. (2004, 2005), which will allow us
 to obtain a more reliable and less
model-dependent estimate of their clustering properties than given in RDA03.

The angular correlation function, $\omega(\theta)$, is a measure of the
clustering of galaxies as projected on the plane of the sky. It is related to
the {\it intrinsic} clustering in three dimensions, parameterised by the
correlation radius $r_0$, by means of an integration known as Limber's formula
(see e.g. Efstathiou et al. 1991, RDA03)

If the two-point correlation function, expressed in proper co-ordinates,
$\xi(r)$ is represented by the simple
model
  $$\xi(r,z)=(r/r_0)^{-\gamma}(1+z)^{-(3+\epsilon)}$$
where
$\gamma\simeq 1.8$ (observationally) and
$\epsilon$ represents the clustering evolution ($\epsilon=0$ is stable and
$\epsilon=-1.2$ is comoving clustering), then Limber's formula gives
$\omega(\theta)=A_{\omega}\theta^{-(\gamma-1)}$, where
$$A_{\omega}=C_{\gamma} r_0^{\gamma} \int_0^{\infty}
{ (1+z)^{\gamma-(3+\epsilon)}\over
x^{\gamma-1}(z){dx(z)\over dz}}[(N(z)^2] dz/[\int_0^{\infty}N(z)dz]^2$$
where $x(z)$ is the proper distance
and $C_{\gamma}=3.679$ for $\gamma=1.8$.

Hence in order to estimate $r_0$
from the observable $A_{\omega}$,a model or observed $N(z)$ is required, and
for deep samples, the $r_0$  will also depend on $\epsilon$.
However, we can also express the clustering measured at any redshift in terms of a comoving
correlation radius $r_{c0}$, without any assumptions about $\epsilon$, as in comoving space
$$\xi(r_{c},z)=(r_c/r_{c0})^{-\gamma}$$
In this system comoving clustering would be a constant $r_{c0}$ and stable clustering would be
 $r_{c0}\propto(1+z)^{-{2\over 3}}$.

Previously (RDA03) we calculated the angular
correlation function, $\omega(\theta)$ of the CDFS ERGs.We
obtained a $\sim 3\sigma$ detection of clustering at the completeness limit
of $K_{s}\leq 21.5$, which
appeared consistent with
the ERG $\omega(\theta)$ measurements  of
Daddi et al. (2000), Firth et al. (2002) and
Roche et al. (2002). By combining all these
we estimated $r_{c0}=12.5\pm 1.4 h^{-1}$ Mpc for ERGs.
However, this estimate is dependent on a model $N(z)$, taken from a `merging and
negative density evolution' model, fitted to
the ERG number counts without any redshift data.
 Furthermore the clustering evolution remains undetermined.

We can now obtain  a less model-dependent
 estimate of  $r_{c0}$, and investigate $\epsilon$,
  by including redshift data.
 We assign to each ERG a redshift
estimate, which for  the 16   galaxies in our redshift sample is the GMOS
spectroscopic redshift
and for the remainder, the photometric redshift
estimate from Caputi et al. (2004, 2005).
Objects with $z_{phot}>4$
 are assigned redshifts following Section 4.6 of
 Caputi et al. (2004), in which only two ERGs with
 $K_{s}<22$ (one with $K_{s}<21.5$) are
 accepted as probable $z>4$ galaxies, and four of our original sample are
reclassified as stars.

     Figure 7 shows the distribution of
true/estimated redshifts for the 175 ERGs  brighter than the approximate
 completeness  limit $K_{s}=21.5$. The photometric $N(z)$ is quite close to
the RDA03 model, especially at the peak, although more dispersed to
lower and higher redshifts, presumably as real galaxies are more
varied in their evolution.
\begin{figure}
\psfig{file=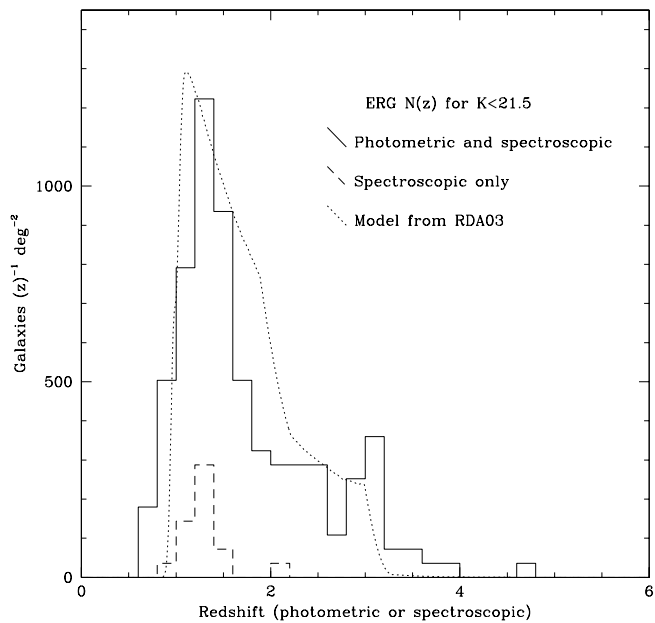,width=90mm}
\caption{The distribution $N(z)$ of photometrically estimated
 (Caputi et al. 2004, 2005), or spectroscopic (this paper)
 redshifts for the 175 $K_{s}\leq 21.5$
ERGs in our CDFS sample (solid histogram); spectroscopic redshifts only
 (dashed histogram) and the best-fit model ERG $N(z)$ from RDA03 (dotted
 line).}
\end{figure}

       The angular correlation functions $\omega(\theta)$ were calculated for
the  CDFS ERGs using the same
methods as in RDA03. Each $\omega(\theta)$
 was fitted with the function `$A_\omega(\theta^{-0.8}-12.24)$'
(see RDA03), to obtain a
      power-law amplitude, $A_{\omega}$.
The errors on the $A_{\omega}$ were estimated using the same
 `bootstrap' method as RDA03
(essentially repeating the analysis for a series of data
catalogs with small regions excluded and finding the scatter in the
resulting estimates).

 We first consider the full $K_s<21.5$ sample of 175 ERGs. From its $\omega(\theta)$
amplitude measurement (Table 6) and photometric/spectroscopic $N(z)$ (Figure 7), Limber's formula gives
 the comoving correlation radius $r_{c0}=13.13^{+2.12}_{-2.44}\rm h^{-1}$ Mpc.
This is consistent with the RDA03 estimate based on a model $N(z)$.
Alternatively if we set $\epsilon=0$
 this result can be fit with a stable clustering model with $r_0=23.91^{+3.86}_{-4.44}\rm h^{-1}$ Mpc
 (at $z=0$).

We then recalculate $\omega(\theta)$ for the ERGs divided into two subsamples in phot/spec redshift.
 For both the low and high redshift subsamples, the $\omega(\theta)$ and $N(z)$ are used to derive a
 comoving $r_{c0}$ (Table 6). Dividing the sample reduces the significance of
clustering detection, but in most cases it remained $>2\sigma$.
We repeat this analysis for redshift divides at $z=1.4$, 1.6 and 1.8.

The subset $\omega(\theta)$ amplitudes
are given in Table 6 and plotted on Figure 8.
It can immediately be seen that the mean of the $r_{c0}$ we derive for the
low and high-redshift subsamples is a little lower than that of the full $K$-limited sample, by about
19 per cent, which corresponds to 32 per cent in the $A_{\omega}$.  The reason for this is that photometric redshifts are not always accurate and thus the low and high
 redshift subsamples will both experience some contamination from galaxies in the other redshift range,
  which are not correlated with the correctly selected galaxies and dilute the observable clustering.
  If, for example, the true $\omega(\theta)$ of the lower redshift ERGs is $A_{l}$ and of the high $A_{h}$,
   and we examine a low $z_{phot}$ subsample with a contamination $\alpha$, the $\omega(\theta)$ we calculate
   will be $A_{obs}\simeq (1-\alpha)^2 A_{l} + \alpha^2 A_{h}$. The observed shortfall could be accounted
   for by $\alpha\simeq 0.20$.
  We have not attempted to correct for this is deriving a $r_{c0}$, and thus the low and high redshift
  $r_{c0}$ will both be slight underestimates, (by similar amounts, as here
  the two subsamples have similar size and clustering).

  Secondly, it can be seem that the higher redshift subsamples have a comoving $r_0$  similar to or
  even a little greater than the lower redshift ERGs. This is clearly more consistent with comoving
  clustering rather than stable clustering. The slight increase in $r_{c0}$ with redshift, if real
  could reflect a stronger clustering for more massive/luminous ERGs.
\begin{table}
\caption{Angular correlation function ($\omega(\theta)$) amplitudes
(given at $\theta=1$ deg. for a fitted $\theta^{-0.8}$ power-law),
estimated for CDFS ERGs, to a limit $K_{s}=21.5$
(Vega system), divided into two subsamples in
 photometric/spectroscopic redshift. This is shown
  for redshift divides $z=1.4, 1.6 and 1.8$}
\begin{tabular}{lcccc}
\hline
$z_{phot/spec}$ & $z_{mean}$ & No. of & $\omega(\theta)$ amp. & $r_{c0}$\\
\smallskip
range &   & ERGs & $10^{-3}(\rm deg^{0.8})$ &  $h^{-1}$ Mpc \\
All & 1.742 & 175 & $6.54\pm 2.02$ & $13.31^{+2.12}_{-2.44}$\\
$z<1.4$ & 1.125 & 72 & $10.94\pm 3.89$ & $9.63^{+1.78}_{-2.08}$ \\
$z>1.4$ & 2.187 & 103 & $7.39\pm4.22$ & $11.39^{+3.25}_{-4.27}$ \\
$z<1.6$ & 1.231 & 101 & $6.61\pm 2.93$ & $8.45^{+1.91}_{-2.35}$ \\
$z>1.6$ & 2.458 & 74 & $8.34\pm 3.46$ & $12.32^{+2.62}_{-3.17}$ \\
$z<1.8$ & 1.273 & 111 & $6.42\pm 2.89$ & $9.03^{+2.03}_{-2.56}$ \\
$z>1.8$ & 2.577 & 64 & $12.41\pm 4.70$ & $14.04^{+2.74}_{-3.26}$ \\
\hline
\end{tabular}
\end{table}
\begin{figure}
\psfig{file=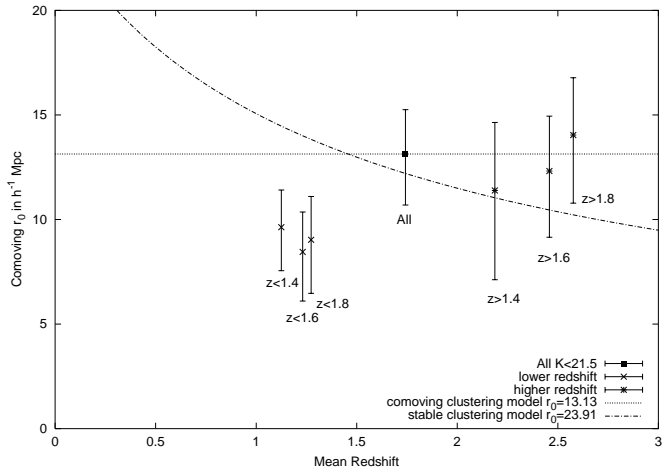,width=90mm,angle=-90}
\caption{Comoving correlation radius $r_{c0}$, derived from the
       $\omega(\theta)$ amplitude and photometric $N(z)$,for a
       full  $K_s<21.5$ sample of ERGs, and for these ERGs divided by photo-$z$
       into low and high redshift
       subsamples (dividing at $z=1.4$, 1.6 and 1.8) plotted against sample mean
       redshift.Also plotted are comoving ($\epsilon=-2$) and stable-clustering ($\epsilon=0$) models
       normalized to the full sample's clustering.}
\end{figure}
Comoving clustering of
$r_0\simeq 12$--$13 h^{-1}$ Mpc is consistent with the clustering of both
shallower (Daddi et al. 2000) and deeper ERG samples, e.g. Daddi et al
(2004) estimate $r_0=9$--$17 h^{-1}$ Mpc at $1.7<z<2.3$.
Extrapolated to lower redshifts,
this model would overpredict the clustering of local passive
galaxies of all luminosities, $r_0=7.2 h^{-1}$ Mpc
(Madgwick et al. 2003), but be consistent with the clustering
of the most luminous ($M_R<-22.27$) early-types at $0.3<z<0.9$,
$r_0=11.2\pm 1.0 h^{-1}$ Mpc
comoving (Brown et al. 2003). This can be explained if (i) many of the
$z>1$ ERGs undergo further merging to become very  massive
ellipticals, (ii)  only some ($\sim $ half) of the moderate-luminosity E/S0s found locally have
evolved directly
from $z>1$ ERGs
and the others
 are formed at $z\leq 1$
from mergers of
less strongly clustered spirals, thus diluting the clustering.

To better understand the evolution of ERGs it will be useful to
compare their clustering with that of their
their likely progenitors at higher redshifts, such as
sub-mm galaxies, for which the clustering
 will be measured in  the ongoing SHADES survey.
There is already evidence that the most massive Lyman break galaxies at
$z\sim 4$ have a similar comoving $ r_0$ ($11.4\pm 2 h^{-1}$ Mpc)
   to the $z\sim 1$--2 ERGs (Allen et al. 2005).

\section{Discussion}

\subsection{Ages of ERGs}
We perform spectroscopic age-dating of the ERGs by fitting their
SEDs with passively evolving models (from Jimenez et al. 2004), and
obtain mean stellar ages ranging widely from 0.6 to 4.5
Gyr, with a mean of $2.1\pm 0.3$ Gyr.
We can compare these results with similar age-dating analyses performed
by other authors, though must be careful to consider the differences in
 how the sample are selected. Firstly
Daddi et al. (2004) give ages (from FORS2 spectra)
for 9 $K$-selected ($K_s<20$) very massive star-forming
galaxies at $1.7<z<2.3$ -- note that not all of these are red enough to be
called ERGs.
Daddi et al. (2005) give ages from low-resolution
ACS spectra of 7 ERGs at $1.39<z<2.47$, of which 5 are passive ellipticals
and 2 weakly star-forming.  McCarthy et
al. (2004) using Gemini Deep Deep Survey
 spectra, give ages for a varied sample of
  20 red ($I-K>3.5$) galaxies at $1.3<z<2.2$. Finally,
 Longhetti et al. (2005) using low-resolution NIR spectra (observed with NICS on
 the La Palma TNG), estimate mass-weighted ages for 10 of the most
bright ($K^{\prime}=16.6$--18.4) and massive  ERGs ($R-K^{\prime}>5.0$) present at $z\sim 1.5$.

Figure 9 plots
stellar age $T_{pas}$ against spectroscopic redshift for all these
 samples,  with loci of the redshifts corresponding to time
$T_{pas}$ before observation.
\begin{figure}
\psfig{file=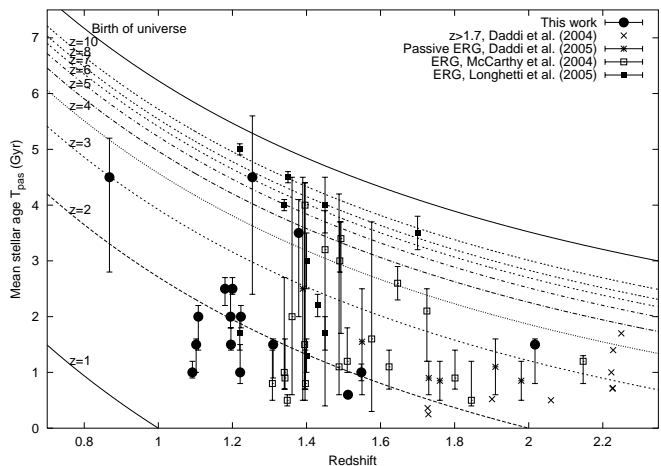,width=90mm,angle=-90}
\caption{Mean stellar age $T_{pas}$, derived from model-fitting to spectra,
  against redshift for our ERG sample, the $K$-selected galaxies of
  Daddi et al. (2004), and the ERGs of Daddi et al. (2005), McCarthy et al. (2004) and
Longhetti et al. (2005).
Plotted loci show the redshifts corresponding to
time $T_{pas}$ before the redshift of observation.}
\end{figure}

For the first four samples the mean stellar ages of ERGs at all $1<z<2.5$
 are spread over the range
between $z\sim 5$
and shortly (a few $\times 10^8$ yr)
 before the epoch of observation. They seem to divide into two groups with
 $z_{msf}\sim2$  and $z_{msf}\sim 3$--4.

The Longhetti et al. (2005) sample of very luminous ERGs differs from the others
in that it
 seems to be concentrated towards higher formation redshifts --
 it contains 5 galaxies with $z_{msf}>5$. These have stellar masses
  3.8--$9.4\times 10^{11}M_{\odot}$, more than twice the mass of
  any ERG in our sample. Hence their ages
  may be evidence that the very most massive spheroidals formed significantly
earlier and/or formed their stars
 most quickly, as in the model of Granato et al. (2004).

 It is also notable that the formation redshifts appear to trace the plotted
 $z_{msf}$ loci. The ages of the oldest galaxies at each redshift, as a function of
 redshift, is an important tracer of the expansion of the universe and hence the
 cosmological model (e.g. Jimenez and Loeb 2002). We would expect
$z_{msf}$ for the oldest galaxies to be constant with observed redshift,
  and the agreement with the plotted loci would support a $\Lambda\simeq 0.75$
 cosmological constant ($w=-1$) model.

As $T_{pas}$ represents a flux-weighted mean stellar age, the true
age of a galaxy
 since formation must always
be greater, to a degree dependent on the detailed form of the
star-formation history.
  One interpretation may therefore be
that most or all of the red galaxies observed as ERGs at $z=1$--2
are galaxies, or mergers of galaxies, which
formed at
 $z\geq 4$--5,  and that the spread in their
stellar ages up to this maximum redshift can be attributed to differences in
 their individual evolution, i.e. the mergers and starbursts they have
 experienced.

Caputi et al. (2004, 2005), using photometric redshift estimates derived
from seven-band photometry of ERGs and other $K<22$ galaxies in
 this field, found the
comoving number density of massive ($>5\times 10^{10}\rm
M_{\odot}$) galaxies in a full $K$-selected sample to fall only slowly from
$z=1.75$ to $z=3.5$, where a significant fraction, 20--25 per cent,
of today's massive galaxies were already in place.
 However, the comoving number density of the reddest subset of these galaxies,
 the ERGs, evolves much more rapidly,
steadily increasing with time from $z=3.5$ to $z=1$, and following a trend
which extrapolates to the present-day E/SO population at $z=0$.
At $z=1$--1.5, the rising comoving number density of ERGs
approached that of all $K$-selected galaxies at the
 higher redshift of $z\sim 3.5$.

Hence, this suggests that a substantial population of massive galaxies
 formed at $z\geq 4$ and were
 initially
starbursting and relatively
blue, but with increasing age many became red enough
to be classed as ERGs.
Interestingly, for the brightest ($I<24.5$) Lyman break galaxies at $z\sim 4$,
Allen et al. (2005) measure strong clustering of $r_0=11.4\pm
 2 \rm h_{100}^{-1}$ Mpc comoving,
consistent with these evolving into the ERGs at $z=1$--2.

To produce the evolution in ERG number density these galaxies must
have entered the ERG class over a wide redshift range
 $1<z<4$, again implying much variation in their star-formation histories.
The formation of so many massive galaxies
at $z\geq 4$ places significant constraints on models of galaxy
formation (e.g. Fontana et al. 2004) and may favour
models with feedback between the star-formation in formative spheroidals
 and the growth of supermassive black holes within them
(Granato et al. 2004; Silva et al. 2005).

\subsection{Ongoing Star-formation Activity in ERGs}
Yan et
al. (2004a) and Doherty et al. (2005),
found that $\geq 75$ per cent of  ERG spectra have prominent absorption features and
$4000\rm \AA$ breaks, while $>50$ per cent also show some [OII] emission.
We find the same for our slightly deeper sample.
For only 3 ERGs do we find neither [OII] emission nor any
indication of a young stellar population from a blue excess in the SED. This is
consistent
 with the
Doherty et al. (2005) estimate
that 28 per cent of ERGs show no evidence of recent
star-formation.

For a clear majority of ERGs, 10/13 at $z<1.5$,
we detect a $\rm [OII]3727 \AA$ emission line.
This includes two galaxies where
the line only just detected with an equivalent width $\sim 5\rm \AA$,
while at the other extreme we have
3 ERGs with [OII] equivalent widths $\geq \rm 30\AA$.
For the 10 emission-line galaxies, the $\rm[OII]3727 \AA$
fluxes - uncorrected for dust - correspond to a mean
SFR
 of 1.63 $\rm M_{\odot} yr^{-1}$, intermediate between the
 2.8 $\rm M_{\odot} yr^{-1}$ mean $\rm SFR_{OII}$ of Yan et al. (2004a) and
the 1.1 $\rm M_{\odot}  yr^{-1}$ of the Doherty et al. (2005) ERGs.

Of course the true SFRs of the emission-line ERGs
will be higher. On the basis of the dust extinction estimated from our model fits, we estimate
the real mean SFR will be about 12--24 $\rm M_{\odot}  yr^{-1}$. This may still be an underestimate as
emission line regions of current SF may suffer more extinction than the continuum produced by
SF over a longer period.
Spitzer mid-IR observations (Yan et al. 2004b) of ERGs similarly
 selected to those in our sample indicate that about
 half have $\rm SFR>12 M_{\odot} yr^{-1}$, and for these
 the mean SFR $\simeq 60 \rm M_{\odot}\rm yr^{-1}$.
We should be able to obtain genuinely dust-independent measurements of the SFRs
for the individual ERGs in the GOODS-S from the  deep Spitzer
$24\rm \mu m$ survey of this field scheduled for later this year.

\subsection{Morphology and Star-formation History}
Examining the HST-ACS images of ERGs we find ERGs to be a mixture of spheroidals, spirals and merging systems.
 The classifications by eye are confirmed by Sersic model fits to the radial profiles (Table 5).
We discuss the nature and SF activity of the ERGs within each of these morphological
categories.

\subsubsection{Spheroidals}
We class 7 ERGs, almost half the sample, as spheroidals. Spectroscopically
they are varied, and this may be related to whether or not they have recently undergone
a merger event or the accretion of a gas-rich galaxy. One spheroidal (2401) appears to be old and purely
passive, with no evidence of SF in the last 2.5 Gyr. Another (1955) shows no
 [OII] emission but its near-IR colours suggest it may be a
  dust-reddened post-starburst galaxy.
One of the spheroidal ERGs hosts an obscured AGN, it may also be star-forming. One is old with a young,
 dusty starburst (1143). The final three (2246, 1152, 1029) are relatively young
  (1--2 Gyr) with the model fits suggesting
prolonged, low-level, non-dusty SF; and appear to show post-merging features.

\subsubsection{Mergers}
Of these 16 ERGs 5 are obvious mergers, with multiple nuclei,
and all of these are star-forming galaxies, on the basis of [OII] emission.
 The two $z>1.5$ mergers also have a high
 surface brightness (Figure 6). Presumably the SFRs and dust
 content are both increased by the interactions.
Some appear to be mixed mergers
 of passive spheroidals with
bluer irregulars  (e.g. 1565 and 1823), and
one (1762) has prominent tidal tails. On the basis of our model fits,
 four of the 5 mergers are relatively young
 (0.6--2.0 Gyr) and undergoing short-term (10--40 Myr) dusty starbursts. Notably,
  in our sample; 4 of the 5 galaxies
 with $T_{sb}<50$ Myr are mergers (the other is an AGN). This would be expected from the current interaction
 acting to tidally
 trigger an intense but short-term starburst. The fifth merger, a close irregular/spheroidal pair,
 (1823) differs in that it is $\sim 4.5$ Gyr
  old with a long $T_{sb}$ and Balmer absorption. This may be explained if
   it is a  late-stage merger where most
  starbursting occurred at earlier encounters of the two galaxies, and the elliptical is already very old.

\subsubsection{Spirals}

Perhaps a more surprising discovery is that of a third
 morphological class of ERGs which are neither ellipticals nor mergers,
 but red spirals. We find four of these.
One explanation for the red colours of these galaxies
 is that they are very dusty, the ERG
 colour criterion selecting out the spirals at the very top end of the range
 of internal dust extinctions, the other is that they long ago ceased forming
 stars. According to our model, the four spirals differ widely in age (1.0--3.5 Gyr)
  but in all cases contain a young component with a long $T_{sb}$ (200--800 Myr)
  and heavy dust extinction ($E(B-V)\simeq 0.5$).
   Three show [OII] emission, the fourth (1777) may therefore have ceased
  forming stars. Strong Balmer lines are  seen in two of the spirals, 1412 and 1876, and this reinforces
   the evidence from model fits that spiral ERGs
    tend to contain large intermediate-age (0.1--1.0 Gyr) populations.

 It is also notable that spiral
1876 has a luminous red (passive?) nucleus
 ringed by blue star-forming regions, producing a strong colour gradient.
 Indeed, for 3 of the spiral ERGs, a comparison of the $I$ and $B$ images shows that
  (in contrast to the merging ERGs), their star-formation is
 not predominantly nuclear but in the outer disks.
  The $z=1.55$ ERG
\#4950 of Daddi et al. (2005), classed as type Sa,
also has a red nucleus with a blue outer ring.

The existence and spectral properties of very red spirals can be explained if
they are galaxies with continuous star-formation which has been
gradually slowed or completely truncated. This explains the
long $T_{sb}$, Balmer absorption, and the absence of emission lines in one spiral (already truncated?).
In addition to this, they are very dusty, as the ERG
 colour criterion would select out the spirals at the very top end of the wide range
 of internal dust extinctions.

Very red spiral galaxies also
 occur at lower redshifts, where they make up some 0.3 per cent of
 all galaxies (Yamauchi and Goto 2004), and tend to be
found in the outer regions of clusters (Couch et al. 2002;
 Goto et al. 2003).
In the model of Bekki, Couch  and Shioya (2002)
a spiral galaxy, infalling into a cluster, experiences a
truncation of star-formation activity, as its gas content is removed
 through ram-pressure stripping by the intra-cluster medium
 and/or by cluster tidal  fields.
Following this the galaxy would become very red
in $\sim 1$ Gyr but could retain a spiral appearence, with visible arms,
  for $\geq 3$ Gyr, although would ultimately
evolve into an S0 type.
This mechanism could give rise to red spirals at ERG redshifts, as
both spiral galaxies (Dawson et al. 2003) and large ERG-rich
clusters (e.g. Kurk et al. 2004)
are already in place at $z\geq 2$.
 However, some spiral ERGs e.g. 1876, must be in relatively early stages of this process
  as much SF is still ongoing in their outer regions.

More information on the nature of spiral ERGs and the relative importance of
dust and `truncation' will soon be available from Spitzer $24\mu \rm m$ data,
giving a dust-independent SFR, and
spectroscopy at
 1--$2\rm \mu m$, showing Balmer lines and $H\alpha$.
  It is likely that some sort of
  environmental effect is involved because ERGs in general are very much more
clustered than typical spirals.

\subsection{Further investigations}

Ongoing surveys of the CDFS with the Spitzer infra-red
telescope will, when complete,
 provide accurate, dust-independent measurements of the
SFRs in our redshift sample.
The ongoing SHADES survey of bright,
$F(850\rm \mu m)>8 mJy$,
 sub-mm galaxies will
provide a measurement
 of the clustering of these probable progenitors of the ERGs (Dunlop 2005;
 Mortier et al. 2005).
 We have planned further and more extensive spectroscopic surveys,
including in the near-IR, of high redshift galaxy clusters and large-scale
 structure.

\section*{Acknowledgements}
This paper is based on observations obtained at the Gemini Observatory, which
is operated by the Association of Universities for Research in Astronomy on
behalf of the Gemini partnership. This paper also makes use of publically
available GOODS data from the
Advanced Camera for Surveys on the NASA Hubble Space Telescope and from the
Infrared Spectrometer and Array Camera on the Very Large Telescope operated by
the European Southern Observatory. We thank the GOODS team for
providing reduced data products. JSD and NDR acknowledge PPARC funding.
NDR acknowledges funding from the Universidad Nacional Autonoma de Mexico.
\section*{References:}
\vskip0.15cm \noindent Abraham R.G., Glazebrook K., McCarthy P.J., Crampton D., Murowinski R., Jorgensen I., Roth K., Hook I. M., Savaglio S., Chen, H.-W.,
 Marzke R. O., Carlberg R.G., 2004, AJ, 127, 2445.

\vskip0.15cm \noindent Alexander D.M., Vignali C., Bauer F.E., Brandt W. N.,
Hornschemeier . E., Garmire G.P., Schneider D.P., 2002, AJ, 123, 1149.

\vskip0.15cm \noindent Allen P.D., Moustakas L.A., Dalton G., MacDonald E.,
 Blake C., Clewley L., Heymans C., Wegner G., 2005, MNRAS, in press.

\vskip0.15cm \noindent Afonso J., Georgakakis A., Hopkins A.M., Sullivan M.,
Mobasher B., 2005, ApJ, submitted.

\vskip0.15cm \noindent Bekki K., Couch W. and Shioya Y. 2002, ApJ, 577, 651

\vskip0.15cm \noindent Brown, M.J.I., Dey A., Jannuzi B.T., Lauer T. R., Tiede G.P., Mikles V.J.,2003, ApJ, 597, 225.

\vskip0.15cm \noindent Bruzual G., Charlot S., 2003, MNRAS, 344, 1000.

\vskip0.15cm \noindent Calzetti D., Armus L., Bohlin R.C., Kinney A.L.,
Koornneef J., Storchi-Bergmann T., 2000, ApJ, 533, 682.

\vskip0.15cm \noindent Caputi K.I., Dunlop J.S., McLure R.J., Roche N.D.,
2004, MNRAS, 353, 30.

\vskip0.15cm \noindent Caputi K.I., Dunlop J.S., McLure R.J., Roche N.D.,
2005, MNRAS, in press.

\vskip0.15cm \noindent Cimatti A., et al., 2002, A\&A, 381, 68.

\vskip0.15cm \noindent Cimatti A., et al., 2003, A\&A, 412, L1.

\vskip0.15cm \noindent Couch W.J., Balogh M.L., Bower R.G., Smail I.,
	Glazebrook K., Taylor M., 2001, ApJ, 549, 820.

\vskip0.15cm \noindent Daddi E., Cimatti A., Pozzetti L., Hoekstra H.,
R{\"o}ttgering, H., Renzini A., Zamorani G., Mannucci F, 2000, A\&A, 361, 535.

\vskip0.15cm \noindent Daddi E., et al,  2004, ApJ., 600, L127.

\vskip0.15cm \noindent Daddi E., et al., 2005, ApJ, 626, 680.

\vskip0.15cm \noindent Dawson S., McCrady N., Stern D., Eckart M.E.,
Spinrad H., Liu M.C, Graham J.R., 2003, AJ, 125, 1236.

\vskip0.15cm \noindent Doherty M., Bunker A.J., Ellis R.S., McCarthy P.J.,
2005, MNRAS, 361, 525.

\vskip0.15cm \noindent Driver S.P., Liske J., Cross N.J.G., De Propris R.,
Allen P.D., 2005, MNRAS, 360,81.

\vskip0.15cm \noindent Dunlop J.S., Peacock J., Spinrad H., Dey A., Jimenez R., Stern D., Windhorst R., 1996, Nature, 381, 581.

\vskip0.15cm \noindent Dunlop J.S., `Starbursts: From 30 Doradus to
Lyman Break Galaxies', Astrophysics \& Space Science Library, Vol. 329.
Dordrecht: Springer, 2005, p.121

\vskip0.15cm \noindent Efstathiou G., Bernstein G., Katz N., Tyson J.A.,
Guhathakurta P., 1991, ApJ, 380, L47.

\vskip0.15cm \noindent Firth A.E., et al., 2002, MNRAS, 332, 617.

\vskip0.15cm \noindent Floyd, D.J.E., Kukula, M.J., Dunlop J.S.,
McLure R.J., Miller L., Percival W.J., Baum S.A.,  O'Dea C.P., 2004,
 MNRAS, 355, 196.

\vskip0.15cm \noindent Fontana A., et al., 2004, A\&A, 424, 23.

\vskip0.15cm \noindent Giacconi R., et al., 2002, ApJS, 139, 369.

\vskip0.15cm \noindent Giavalisco et al., 2004, ApJ, 600, L93.

\vskip0.15cm \noindent Goto T., et al, 2003, PASJ, 55, 757.

\vskip0.15cm \noindent Granato G.L, De Zotti G., Silva L., Bressan A., Danese,
L., 2004, ApJ, 600, 580.

\vskip0.15cm \noindent Hamuy M., Suntzeff N., Heathcote S., Walker A., Gigoux
P., Phillips M., 1994, PASP, 106, 700, 566.

\vskip0.15cm \noindent Jimenez R., Loeb A., 2002, ApJ, 573, 37.

\vskip0.15cm \noindent Jimenez R., MacDonald J., Dunlop J.S., Padoan P.,
Peacock J.A., 2004, MNRAS, 349, 240.

\vskip0.15cm \noindent Kennicutt R., 1998, ARA\&A, 36, 189.

\vskip0.15cm \noindent Kurk J.D., Pentericci L., R\``{o}ttgering H.J.A., Miley G.K., 2004, A\&A, 428, 793.

\vskip0.15cm \noindent Longhetti M., et al., 2005, MNRAS, 361, 897.

\vskip0.15cm \noindent McCarthy P.J., et al., 2004, ApJ, 614, L9.

\vskip0.15cm \noindent Madgwick D.S., et al., 2003, MNRAS, 344, 847.

\vskip0.15cm \noindent de Mello D., Daddi E., Renzini A., Cimatti A., di
Serego Alighieri A., Pozzetti L., Zamorani G., 2004, ApJ, 608, L29-L32.

\vskip0.15cm \noindent Mortier A.M.J., et al. 2005, MNRAS, 363, 509.

\vskip0.15cm \noindent Roche N.D., Almaini O., Dunlop J.S., Ivison R.J.,
Willott C.J., 2002, MNRAS 337, 1282.

\vskip0.15cm \noindent Roche N.D, Dunlop J.S., Almaini O., 2003, MNRAS, 346, 803. [RDA03]

\vskip0.15cm \noindent Shen S., Mo H.J., White S.D.M., Blanton M.R., Kauffmann
G., Voges W., Brinkmann J., Csabai1 I., 2003, MNRAS, 343, 978.

\vskip0.15cm \noindent Silva L., De Zotti G., Granato G.L., Maiolino R., Danese L., 2005, MNRAS, 357, 1295.

\vskip0.15cm \noindent Spergel D.N., et al, 2006, preprint.

\vskip0.15cm \noindent Szokoly G.P., et al., 2004, ApJS,155, 271.

\vskip0.15cm \noindent Vanzella E., et al., 2005.

\vskip0.15cm \noindent Yamauchi C., Gotu T., 2004, MNRAS, 352, 815.

\vskip0.15cm \noindent Yan L., Thompson D.J, 2003, Ap.J. 586, 765.

\vskip0.15cm \noindent Yan L., Thompson D.J, Soifer B.T., 2004a, AJ, 127, 1274.

\vskip0.15cm \noindent Yan L., et al., 2004b, ApJS,154, 75.

\vskip0.15cm \noindent Zheng W., et al., 2004, ApJS,155, 73.
\end{document}